\documentclass{aa}  
\usepackage{graphicx}
\usepackage{txfonts}
\usepackage{amsmath}
\usepackage{multirow}
\usepackage{subfig}
\usepackage{color}
\usepackage{url}
\usepackage{booktabs}
\usepackage{rotating}
\usepackage{setspace}

\definecolor{dodgerblue}{RGB}{30,144,255} 
\begin{document} 
\newcommand{\ergs}{erg\,s$^{-1}$}
\newcommand{\kev}{keV }
\newcommand{\ra}{r_\mathrm{A}}
\newcommand{\rpl}{r_\mathrm{pl}}
\newcommand{\rns}{R_\mathrm{NS}}

   \title{Pulsating ultraluminous X-ray sources: modeling the thermal emission and polarization properties}


   \author{S. Conforti
          \inst{1,} \inst{2}
          \and
          L. Zampieri \inst{2} 
          \and
          R. Taverna \inst{3}
          \and 
          R. Turolla \inst{3,} \inst{4}
          \and 
          N. Brice \inst{6}
          \and
          F. Pintore \inst{5}
          \and
          G.L. Israel \inst{7}
          }

   \institute{Dipartimento di Fisica e Astronomia ``G. Galilei'', Università degli Studi di Padova, Vicolo dell'Osservatorio 3, 35122 Padova, Italy\\
              \email{silvia.conforti@phd.unipd.it}
         \and
             INAF-Osservatorio Astronomico di Padova, Vicolo dell'Osservatorio 5, 35122 Padova, Italy 
         \and 
             Dipartimento di Fisica e Astronomia ``G.Galilei'', Università degli Studi di Padova, Via Marzolo 8, 35121 Padova, Italy 
         \and 
             Mullard Space Science Laboratory, University College London, Holmbury St. Mary, Surrey, RH5 6NT, UK 
         \and 
             INAF/IASF Palermo, Via Ugo la Malfa 153, 90146 Palermo, Italy
        \and
            Mullard Space Science Laboratory, University College London, Holmbury St Mary, Dorking, Surrey RH5 6NT, UK
        \and
            INAF–Osservatorio Astronomico di Roma, via Frascati 33, I-00078 Monte Porzio Catone, Italy
             }

   \date{Received ...; accepted ...}

 
  \abstract
   {{Ultraluminous X-ray sources (ULXs) are enigmatic sources first discovered in the 1980s in external galaxies. They are characterized by their extraordinarily high X-ray luminosity, which often exceeds $10^{40}\, \rm{erg \; s^{-1}}$}.}
   {{Our study aims to obtain more information about pulsating ULXs (PULXs), first of all, their viewing geometry, since it affects almost all the observables, such as the flux, the pulsed fraction, the polarization degree (PD), and polarization angle (PA).}}
   {{We present a simplified model, which primarily describes the thermal emission from an accreting, highly magnetized neutron star, simulating the contributions of an accretion disk and an accretion envelope surrounding the star magnetosphere, both described by a multicolor blackbody. Numerical calculations are used to determine the flux, PD, and PA of the emitted radiation, considering various viewing geometries. The model predictions are then compared to the observed spectra of two PULXs, M51 ULX-7 and NGC 7793 P13.}}
   {{We identified the best fitting geometries for these sources, obtaining values of the pulsed fraction and the temperature at the inner radius of the disk compatible with those obtained from previous works. We also found that measuring the polarization observables can give considerable additional information on the source.}}
   {}

   \keywords{Neutron Stars --
                X-ray emission --
                Polarization
               }

   \maketitle
%

\section{Introduction}
Ultraluminous X-ray sources (ULXs) were discovered in external galaxies in the 1980s with the Einstein X-ray Observatory \cite[][]{fabbiano1989x,long1981observations}. One of the most distinctive features of ULXs is the high X-ray luminosity, which can reach values of $L\sim10^{41} \rm{erg \; s^{-1}}$, far exceeding the Eddington limit for a solar mass object \cite[see e.g.][for a review]{Kaaret:tesi}. From an observational standpoint, ULXs are characterized by a higher occurrence in star-forming galaxies \cite[][]{Douglas:tesi}, a turnover in the spectra at energies in the 2--10 keV range, a second softer component below 1 keV, and spectral variability between different epochs \cite[e.g.][]{gladstone2009ultraluminous,pintore2012x, sutton2013bright, sutton2013ultraluminous, middleton2015spectral, middleton2015diagnosing}.

The large amount of energy emitted and the X-ray variability led to believe that ULXs were binary systems in which a massive donor transfers material onto a black hole \cite[][]{kaaret2017ultraluminous}.  
For some time, ULXs were thought to be powered by black holes of intermediate mass (IMBHs), that is, larger than $100M_{\odot}$ \cite[][]{colbert1999nature,Makishima:tesi,miller2003x}. In fact, the Eddington luminosity for such objects exceeds the observed one, so accretion can proceed at a sub-Eddington rate. Later on, super-Eddington accretion onto stellar-mass black holes ($M_{BH}\sim 10$--$20\,M_{\odot}$, \citealt{stobbart2006xmm,feng2011ultraluminous} and also black holes with $M_{BH}\sim 30$--$80\,M_{\odot}$, \citealt{zampieri2009low}) were also considered. If this was the case, however, emission needs to be beamed to match the observed luminosity \cite[][]{Fabrika:tesi,King:tesi,Poutanen:tesi}.

This picture was soon revolutionized in 2014, when pulsations were discovered in M82 X-2, a ULX in the M82 galaxy \cite[][]{Bachetti:tesi}, supporting the view that at least some of these sources could be powered by a neutron star (NS). The discovery of the first pulsating ULX was soon followed by several others \cite[][]{furst2016discovery,israel2017accreting,israel2017discovery,castillo2020discovery,Carpano:tesi,Sathyaprakash:tesi,quintin2021new}, highlighting the existence of an entire population of pulsating ULXs, dubbed PULXs. 

Even though the association of PULXs with neutron stars appears well established, explaining how neutron stars can emit such a huge amount of energy remains an open problem, given that the Eddington luminosity for a neutron star is $\approx 10^{38}\,$\ergs.
Accretion onto (highly) magnetized NSs has been the focus of many investigations, starting from the pioneering work of \cite{Basko:tesi}.  
They presented a model for accretion onto magnetized neutron stars that included the formation of an accretion column on the magnetic poles of the star for the first time. This model for the column was somewhat simplistic and did not take into account the consequences of the propeller effect, induced by the huge magnetic field strength ($\sim10^{15}\, \rm{G}$), on the column. Subsequent investigations addressed the problem with an increasing level of sophistication, including the effects of a super-strong magnetic field  \cite[see][]{Lyubarskii:tesi, Mushtukov:tesi,mushtukov2017optically,brice2021super}. In particular, it was realized that in a strongly magnetized accretion column, the opacity is drastically suppressed for photons polarized perpendicularly to the magnetic field direction. This results in a much higher value of the Eddington luminosity, allowing for a larger flux to escape from the sides of the column. Such a scenario could indeed explain the large luminosity observed from these sources.
We present here a simplified model (torusdisk hereafter) that reproduces the X-ray thermal emission of pulsating ULXs, in terms of emission from an accretion disk and an accretion envelope, which is modeled as a torus confined by the magnetic field lines that extend up to the magnetospheric radius. 
To this aim, we adapted the numerical code by \citet{taverna2017spectrum}, which calculates the flux of photons coming from the part in view of the system as a function of both the photon energy and the star rotational phase, so to compute spectra and light curves of PULXs in different energy bands. 
In addition to the flux, the code also computes the polarization degree (PD) and polarization angle (PA) as functions of energy and phase.

Finally, we compare the simulations with the observational data of two PULXs, to test if the model can reproduce the properties exhibited by real sources. Firstly we focus on M51 ULX-7 (aka NGC 5194 X-7, \citealt{roberts2000rosat},  CXOM51 J133001.0+47134, \citealt{terashima2004luminous},  NGC 5194/5 ULX-7, \citealt{liu2005catalogue}), located in the outskirts of a young open cluster in a spiral arm of its host galaxy. It belongs to a high-mass X-ray binary (HMXB) \cite[][]{abolmasov2007spectroscopy} with a (likely) O-B companion with mass $\gtrsim 8\,M_\odot$ \cite[][]{castillo2020discovery, abolmasov2007spectroscopy} and exhibits an X-ray luminosity of $6\times10^{39}$ \ergs. It was suggested that the source is powered by a neutron star accreting at a super-Eddington rate \cite[][]{brightman2022evolution}. 
NGC 7793 P13 (P13 hereafter) is a pulsating ULX located on the southern edge of the galaxy NGC 7793 \cite[in the Sculptor group][]{fabbiano1992x}, the X-ray emission of which is completely dominated by this source. The ESO-VLT observations in 2008--2009 revealed that NGC 7793 P13 has a B9Ia supergiant companion with a mass between $10$ and $20\,M_{\odot}$ \cite[][]{motch2011supergiant,furst2021long}. From {\it XMM-Newton} and {\it NuSTAR} observations in 2016, coherent pulsations were detected ($P\approx 0.42$ s); this led to identify P13 as the third known ULX powered by a neutron star \cite[][]{israel2017discovery,furst2016discovery}.

The main goal of this paper is to derive information on the viewing geometry of these sources confronting the simulated light curves with the observed ones, to constrain the temperature at the inner radius of the disk, and the magnetic field strength. To reproduce the spectrum in the $0.1-10$ keV band, three components are required: the spectrum from our model (torus and disk), a cut-off power law component to account for the non-thermal emission in the hard band (believed to come from the accretion column), and a soft blackbody component, possibly due to radiative winds produced in the intermediate/outer part of the accretion disk \cite[][]{Poutanen:tesi,pinto2016resolved,pinto2020thermal,middleton2022thermally}. In case spectral fitting is degenerate for different geometries, we use polarization to discriminate among them. This further shows that polarimetric measurements are indeed a powerful tool for the study of highly magnetized neutron stars in PULXs, as testified by the results obtained by the recent NASA-ASI mission {\it IXPE} \cite[Imaging X-ray Polarimetry Explorer,][]{weisskopf2022imaging}.

This work is organized as follows. In section \ref{sec:model} we present the model and the assumptions on which it relies. In section \ref{sec:data} we provide a brief description of how the data were analyzed while section \ref{sec:results} is devoted to the spectral and polarization results, which are then discussed in section \ref{sec:discussions}. Finally, our conclusions are summarized in section \ref{sec:summary}.

\section{Physical model}
\label{sec:model}

In this section, we introduce the physical model adopted to simulate and reproduce the X-ray spectrum and polarization properties of the radiation emitted by pulsating ULXs.

\subsection{Modeling the source}
\label{subsec:the source} 
The model computes the emission from an accreting magnetized neutron star (NS) in a binary system. The accreting material originates from the donor star and proceeds through a geometrically-thin accretion disk. At the Alfvén radius ($\ra$),  where the magnetic pressure equals the ram pressure of the gas particles, the matter is funneled along the closed field lines of the magnetic field, falling towards the polar regions. This envelopes the star in a sort of cocoon that, for the sake of simplicity, we call ``torus'' hereafter \cite[][]{Mushtukov:tesi,mushtukov2017optically, brice2021super}. At the high accretion rates expected to occur in PULXs, this torus is optically thick. In the following, we assume that the neutron star has a mass of $1.4\, M_{\odot}$, a radius of $10\,\mathrm{km}$, and that the magnetic field has a dipolar topology with a strength (at the pole) in the range $10^{12}$--$10^{13}\, \mathrm{G}$.
The equation of the magnetic field lines is, in polar coordinates, 
\begin{equation}
    r=r_\mathrm{A}\sin^2{\theta} \, ,
\label{linee di campo}
\end{equation}
where $\theta$ is the magnetic colatitude, and $\ra$, for the sake of simplicity, here coincides with the maximum radius of the magnetosphere, which is computed with the equation (1) in \cite{brice2023observational}:
\begin{equation}
R_{\rm{m}} = 7\times10^{7}\Lambda M^{1/7}R_6^{10/7}B_{\rm{d},12}^{4/7}L_{39}^{-2/7}\,\rm{cm},
\label{alfven_radius}
\end{equation}
where $\Lambda$ is a dimensionless parameter that depends on the mode of accretion (a typical value of 0.5 for accretion via thin disc), $B_{d,12}$ is the surface dipole field strength at the magnetic poles (expressed in units of $10^{12}$ G), and $L_{39}$ is the accretion luminosity (in units of $10^{39}\,\rm{erg\,s^{-1}}$).
The equation (\ref{linee di campo}) is just an approximation since the exact topology of the magnetic field near the surface of the NS is unknown. Pulsating ULXs, in particular, may have a multipolar magnetic field structure, as proposed by \cite{israel2017accreting} to accommodate for the large field required to rise the Eddington limit enough and, at the same time, prevent the star from entering the propeller stage. However, higher-order multipoles quickly decay moving away from the surface, so our assumption is quite justified if $\ra$ is tens stellar radii and makes the computations much simpler.
\begin{figure}
    \centering
    \includegraphics[width=\hsize]{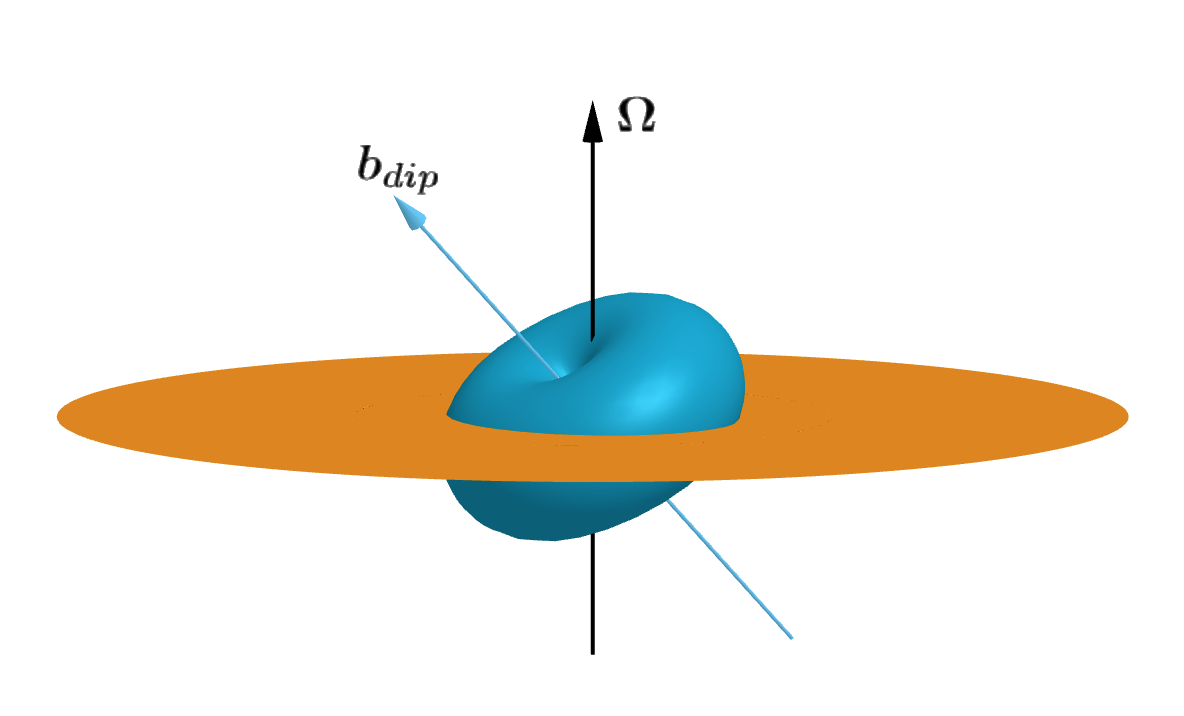}
    \includegraphics[width=\hsize]{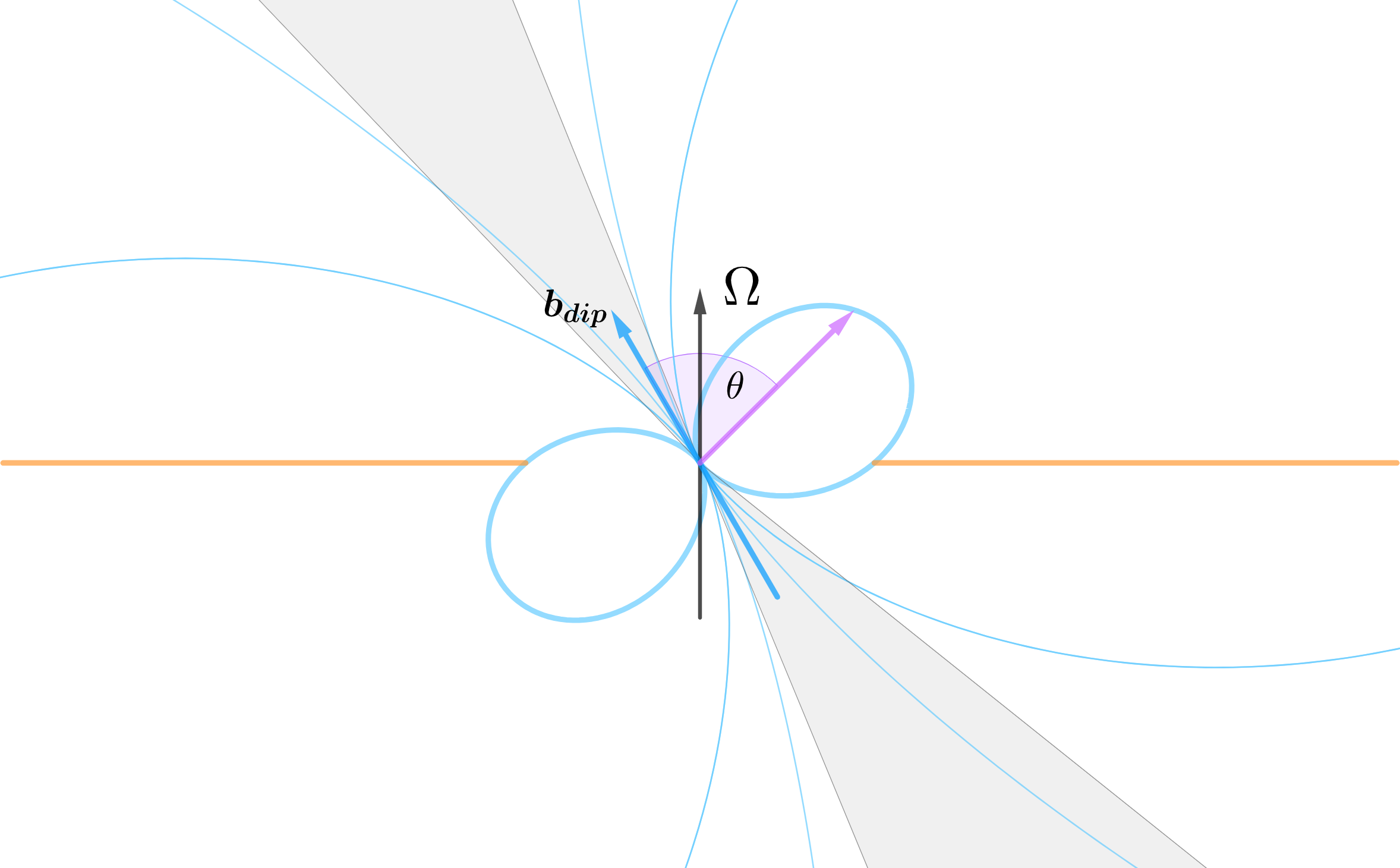}
    \caption{A representation of the system. A 3D view on the top and a meridional cut on the bottom. The accretion disk and torus are shown in orange and blue respectively, the spin axis $\Omega$ and the magnetic dipole axis $b_{dip}$ are also indicated (in gray the radiation emitted from the accretion column).}
    \label{fig:sorgente}
\end{figure}

Both the disk and the torus are assumed to be in Local Thermal Equilibrium (LTE). The disk temperature follows the thin-disc profile \cite[see][]{shakura1973black}
\begin{equation}
    T(r)=T_\mathrm{in} \left( \frac{R_\mathrm{in}}{r} \right)^{3/4},
\label{temperatura del disco}
\end{equation}
where $T_\mathrm{in}$ is the temperature at the inner radius of the disk, $R_\mathrm{in}$, that we take to coincide with the radius at which the torus and the disk intersect. 
The temperature profile of the torus is that presented in \cite{brice2023observational}, it varies with the magnetic colatitude and follows the expression:
\begin{equation}
T_{\rm{out, torus}}=T_{\rm{in, torus}}\tau^{-1/4},
\label{equazione_temp_toro}
\end{equation}
where $T_{\rm{out, torus}}$ is the temperature at the torus surface, $T_{\rm{in, torus}}$ the temperature at the torus inner boundary, and $\tau> 1$ the torus optical depth \cite[see figure 1 in][]{brice2023observational}.
As reported in \cite{brice2023observational}, the local optical depth depends on the dynamics of the accreting particles: the non-uniform velocity field inside the envelope introduces a meridional variation of $T_{\rm{out}}$. Because of our assumption of LTE, each annulus of both the disc and the torus emits blackbody radiation at the local temperature; no effects due to the reprocessing of thermal radiation by material on top of the disc and/or torus are considered. 
In our simplified modeling, we do not account for radiation coming from the sides of the accretion columns which form above the magnetic poles if the mass accretion rate is high enough. Emission from the columns gives rise to a non-thermal tail in the X-ray spectrum (\cite{walton2015broadband},\cite{walton2018super}, \cite{brightman2022evolution}). In order to reproduce the observed spectrum over the entire range, we added to the model a phenomenological cut-off power law, often used to describe this type of non-thermal emission \cite[see e.g.][]{walton2018evidence}

\subsection{Polarization}
\label{subsec:polarization observables}
Radiation emitted by strongly magnetized neutron stars is expected to be highly polarized in two normal modes, the ordinary (O) and extraordinary (X) ones \cite[e.g.][]{gnedin1974transfer, ho2003atmospheres,lai2010polarized}. The former is characterized by an electric field oscillating in the plane of the propagation vector $\boldsymbol{k}$ and the local magnetic field $\boldsymbol{B}$, while in the latter the electric field oscillates perpendicularly to both these vectors. 
The cross-sections of X-mode photons are greatly reduced below the electron cyclotron energy, $E_\mathrm{c,e}\approx 11.6 (B/10^{12}\, \mathrm{G})\,\mathrm{keV}$, by a factor of $\sim (B/B_\mathrm{Q})^2$ with respect to the unmagnetized case, where $B_\mathrm{Q} \simeq 4.414 \times 10^{13}\, \mathrm{G}$ is the critical field \cite[][]{meszaros1992high,harding2006physics}. This makes the medium optically thin for X photons \cite[][]{herold1979compton,ventura1979scattering} and leads to the release of more radiation inside the accretion column, explaining the super-Eddington luminosity in the X-rays.

Moreover, strong magnetic fields are also expected to modify the optical properties of the medium in which radiation propagates. In particular, vacuum birefringence \cite[][]{heisenberg1936folgerungen} affects photons propagating in a strongly magnetized vacuum,  forcing the polarization modes to remain unchanged within a region close to the star surface where the magnetic field is strong enough \cite[][]{heyl2002qed,heyl2003high,fernandez2011x,taverna2014probing}.
Inside the so-called adiabatic region, the scale length $\ell_\mathrm{A}$ along which the photon electric field varies turns out to be much smaller than the scale length $\ell_\mathrm{B}$ that characterizes the variation of the star magnetic field along the photon trajectory. As a result, the photon electric field adapts instantaneously to the local magnetic field direction. On the other hand, far from the star, it is $\ell_\mathrm{A} \gg \ell_\mathrm{B}$, so that the electric field is frozen. The boundary between these two regimes is given by $\ell_\mathrm{A} \simeq \ell_\mathrm{B}$, which is called adiabatic (or polarization-limiting) radius \cite[][]{heyl2003high,taverna2015polarization}: 
\begin{equation}
\frac{\rpl}{\rns}\sim 5 \left( \frac{E}{1\,\mathrm{keV}} \right)^{1/5} \left( \frac{\rns}{10\,\mathrm{km}} \right)^{1/5}  \left( \frac{B_\mathrm{p}}{10^{11}\,\mathrm{G}} \right)^{2/5} \,\,,
\label{adiabatic_radius}
\end{equation}
where $E$ is the photon energy, $B_\mathrm{p}$ the polar magnetic field strength, and $\rns$ the NS radius.
The actual polarization state should be determined by solving the wave equation. However, here we adopt the approximation already discussed in \citet{taverna2015polarization} and assume adiabatic evolution for $r < \rpl$ while the electric field is frozen for $r>\rpl$.    
For the magnetic field strength ($\sim 10^{13}$ G) and the photon energies considered here ($0.1$--$1\, \mathrm{keV}$), the Alfvén radius ($\ra$) is larger than the adiabatic radius ($\ra\sim 50\rns$, while $r_{\rm{pl}} \sim36\rns$), so we expect that a large part of the torus lies outside the adiabatic region. Hence, we assume radiation coming from this zone of the torus to be unpolarized\footnote{This is justified because, whatever the intrinsic polarization is, the polarization degree measured at infinity is close to zero as a consequence of geometric effects \citep[][see also below]{taverna2015polarization}.}. Radiation coming from inside $\rpl$ carries a non-zero intrinsic polarization degree,
\begin{equation}
    \Pi_\mathrm{L}= \left| \frac{N_\mathrm{X}-N_\mathrm{O}}{N_\mathrm{X}+N_\mathrm{O}} \right| \,,
\label{euqzione_pil_intrinseca}
\end{equation}
where $N_\mathrm{X}$ ($N_\mathrm{O}$) is the fraction of X (O) photons. 
Following the same argument, we assume that the polarization of radiation coming from the disk is negligible since the disk starts outside $\rpl$ where $B$ is quite small\footnote{We are interested in the radiation coming from the torus that is polarized in two normal modes, that depend on the magnetic field direction. So, for the sake of simplicity, we neglect the polarization of the disk radiation, since it is different from that of the torus.}.  

It is convenient to express the polarization observables, the polarization degree (PD) and polarization angle (PA), in terms of the Stokes parameters $\mathcal{I}$, $\mathcal{Q}$, and $\mathcal{U}$ \cite[][]{rybicki1991radiative}. In a reference frame with the $z$ axis along the unit wavevector $\vec{k}$, and the $y$ axis in the ($\vec{k},\vec{B}$) plane, the Stokes parameters for X and O photons are 
\begin{equation} 
\begin{pmatrix}
    \mathcal{I} \\  
    \mathcal{Q} \\
    \mathcal{U} 
\end{pmatrix}_X
= 
\begin{pmatrix}
    1 \\
    1 \\
    0
\end{pmatrix}
\,\,\,\,\,\,\,\,\,\,
\begin{pmatrix}
     \mathcal{I} \\  
    \mathcal{Q} \\
    \mathcal{U} 
\end{pmatrix}_O
= 
\begin{pmatrix}
    1 \\
    -1 \\
    0
\end{pmatrix}\, .
\label{stokes_parameters_XO}
\end{equation}
The last Stokes parameter, $\mathcal{V}$, which describes the circular polarization, is not considered here. The Stokes parameters depend on the reference frame of each photon ($x_i, y_i, z_i$), which is fixed by the local direction of the (dipole) magnetic field that changes across the emitting surface. To sum the Stokes parameter we must refer them to the same fixed reference frame, that we choose to be that of the polarimeter. Each photon reference frame is rotated (around $z_i$) with respect to the polarimeter frame by an angle $\alpha_i$. Eventually, for a single photon the Stokes parameters $I_i, Q_i, U_i$ are defined as
\begin{equation}
\begin{aligned}
   I_i &=
   \mathcal{I}_i \\
   Q_i &= 
   \mathcal{Q}_i\cos(2\alpha_i) + \mathcal{U}_i\sin(2\alpha_i) \\
   U_i &=
   \mathcal{U}_i\cos(2\alpha_i) - \mathcal{Q}_i\sin(2\alpha_i).
\end{aligned}
\label{stokes_sin_e_cos}
\end{equation}
The observed PD and PA are given by \cite[][]{rybicki1991radiative}
\begin{equation}
\begin{aligned}
\mathrm{PD} &= 
\frac{\sqrt{Q^2 + U^2}}{I} \\ 
\mathrm{PA} &= 
\frac{1}{2}\arctan{\left( \frac{U}{Q} \right)} \,,
\end{aligned}
\label{pol_frac_and_pol_ang}
\end{equation} 
where the Stokes parameters $Q$ and $U$ refer to the ``total'' radiation, that is, the sum of the $Q_i$ and $U_i$ of the collected photons. Substituting equations (\ref{stokes_parameters_XO}) and (\ref{stokes_sin_e_cos}) inside equations (\ref{pol_frac_and_pol_ang}) one obtains
\begin{equation}
\begin{aligned}
    \begin{split}
    \mathrm{PD} &=\frac{1}{N} \Bigl[ N + 2\sum_i\sum_{k>1}\cos{(2\alpha_i-2\alpha_k)} \\
    &+2\sum_j\sum_{r>j}\cos{(2\alpha_j-2\alpha_r)} \\
    &-2\sum_i\sum_{j}\cos{(2\alpha_i-2\alpha_j)} \Bigr]^{1/2}\\
    \mathrm{PA} &=\frac{1}{2}\arctan{-\frac{\sum_i\sin(2\alpha_i)-\sum_j\sin(2\alpha_j)}{\sum_i\cos(2\alpha_i)-\sum_j\cos(2\alpha_j)}}\,,
    \end{split} 
    \end{aligned}
    \label{sum_pil}
\end{equation}
where $i,k=1,...,N_X$ and $r,j=1,...,N_O$. From equations (\ref{sum_pil}) one can see that the polarization degree measured at infinity is in general lower than that at the emission  (equation \ref{euqzione_pil_intrinseca}), because of the presence of the geometrical factors $\sin(2\alpha)$ and $\cos(2\alpha)$. If the magnetic field topology is particularly tangled (as it occurs close to the star surface), the $\alpha$ angles of the different photons span almost over the entire $0$--$2\pi$ range and the depolarization effect is larger. On the other hand, for a more uniform magnetic field (like that at $r_\mathrm{pl}$, far from the surface), the different $\alpha$ angles attain much more similar values and radiation is much less depolarized \cite[][]{taverna2015polarization}. Hence, when vacuum birefringence is accounted for and the Stokes parameters are computed at $r_\mathrm{pl}$, the polarization pattern at the surface is likely preserved also at the observer.

\subsection{Numerical implementation}
\label{subsec:numerical implementation}
To compute the spectral and polarization properties of the emitted radiation, we used the ray-tracer code discussed in \citet{taverna2015polarization}, but adapted to our model, in particular to a different emitting surface (torus and disk). 

\subsubsection{Visibility of the source}
The code starts by computing the visible part of the source assuming a specific viewing geometry. We consider the surface, comprising the disk and the torus, as a collection of small emitting patches. The grid is made by $50$ bins in $\theta$, $\phi$, and the radial distance $r$, while for both the energy and rotational phase we take 30 bins. The phase is measured from the projection of the line of sight (LOS) onto the plane perpendicular to the spin axis.

\begin{figure}
    \centering
    \includegraphics[scale=0.17]{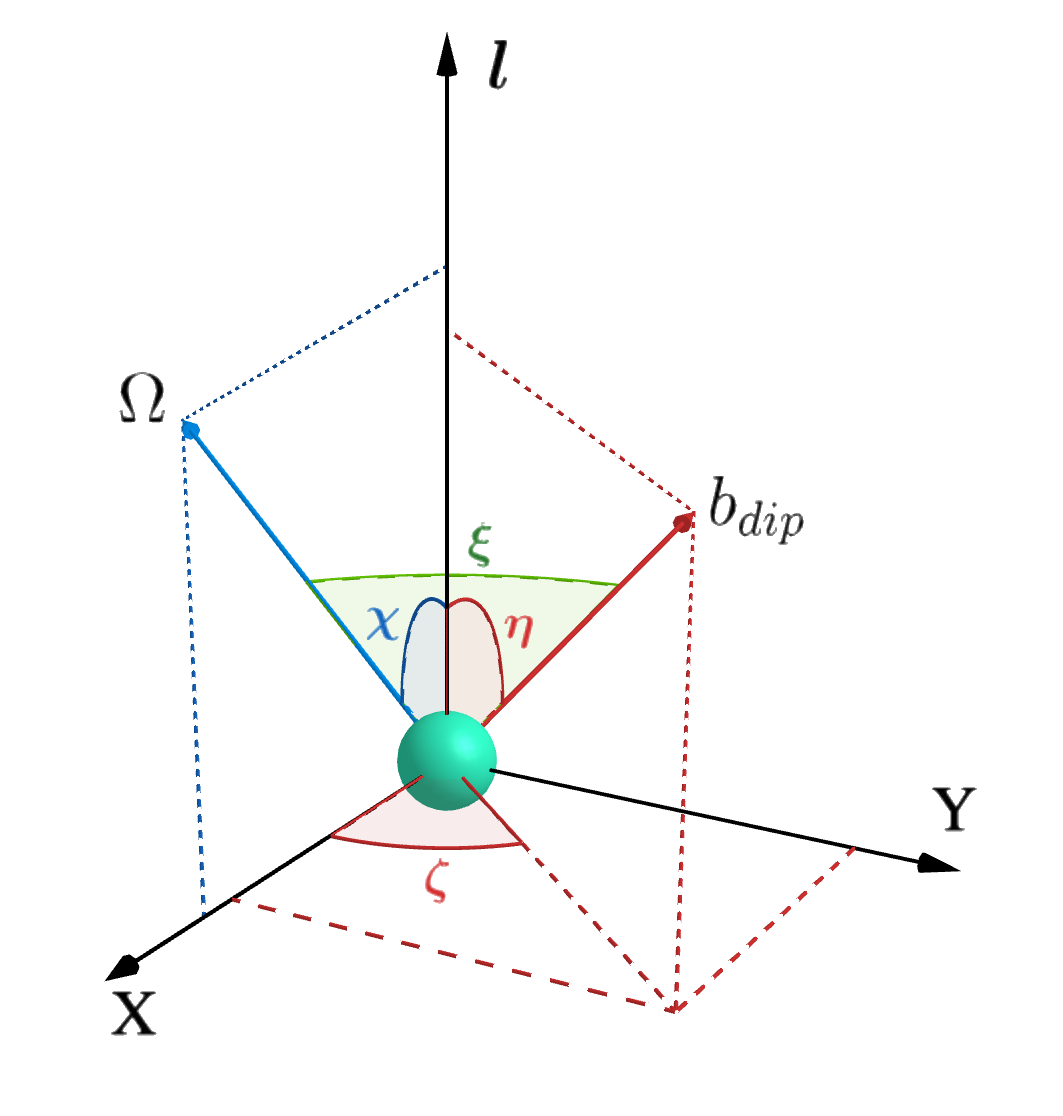}
    \caption{LOS reference frame. The $Z$-axis is aligned to $\boldsymbol{l}$, the LOS unit vector, $\eta$ is the angle between the star magnetic axis $\boldsymbol{b}_\mathrm{dip}$ and the LOS, while $\zeta$ is the corresponding azimuth. The LOS and magnetic axis inclinations $\chi$ and $\xi$ with respect to the spin axis $\boldsymbol{\Omega}$ are also shown.}
    \label{fig:sistema_assi}
\end{figure}

In the LOS reference frame (with the $x$-axis in the plane of $\boldsymbol{l}$ and $\boldsymbol{\Omega}$, see fig. \ref{fig:sistema_assi}), the code selects the points that are in view for the chosen angles $\chi$ and $\xi$, which represent the orientations of the neutron star spin axis and the magnetic axis with respect to the LOS, respectively.
The code also takes into account the orientation of the disk relative to the NS spin axis. The following steps are performed: selecting the visible part of the torus, accounting for its self-shadowing as well as the shadow cast by the disk, and selecting the visible part of the disk not shadowed by the torus.

To achieve this, the code traces a straight line parallel to the LOS and identifies the intersections between this line and the emitting surface (torus and/or disk). To understand if each intersection is in view, we evaluate the quantity 
\begin{equation}
    \boldsymbol{n} \cdot \boldsymbol{l} > 0 \, , 
\label{first condition visibility}
\end{equation}
where $\boldsymbol{n}$ represents the surface normal unit vector at the intersection, and $\boldsymbol{l}$ is the unit vector of the LOS. If multiple intersections satisfy equation (\ref{first condition visibility}), we calculate the $z$-coordinate of each point and consider in view the one with the largest value of the $z$-coordinate. 

\begin{figure*}
\centering
{\subfloat[][\emph{}] {\includegraphics[scale=0.7]{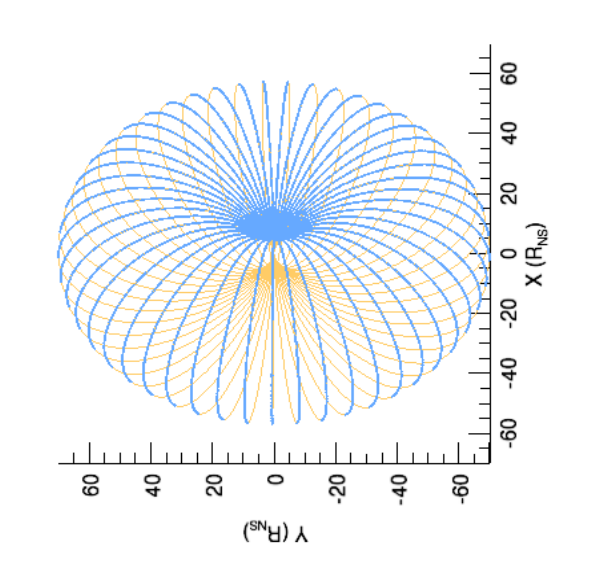}} \quad
\subfloat[][\emph{}] {\includegraphics[scale=0.7]{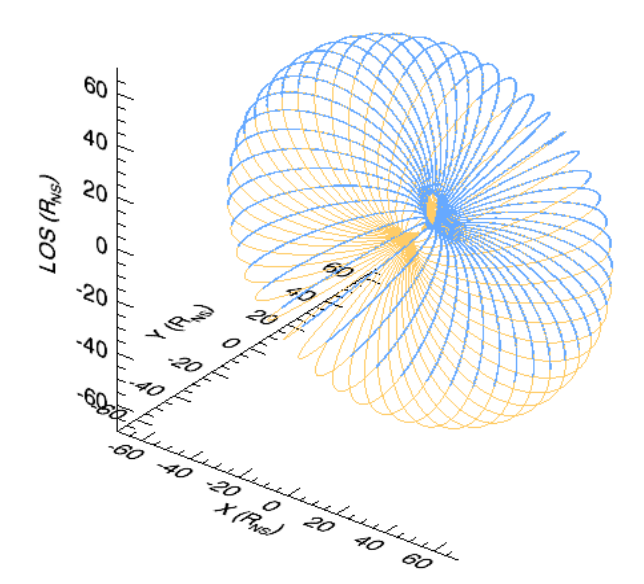}}} \quad
{\subfloat[][\emph{}] {\includegraphics[scale=0.7]{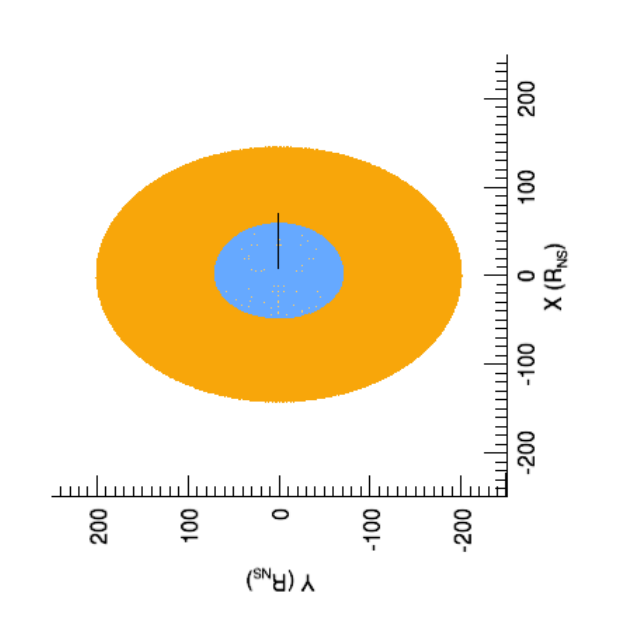}} 
\subfloat[][\emph{}] {\includegraphics[scale=0.7]{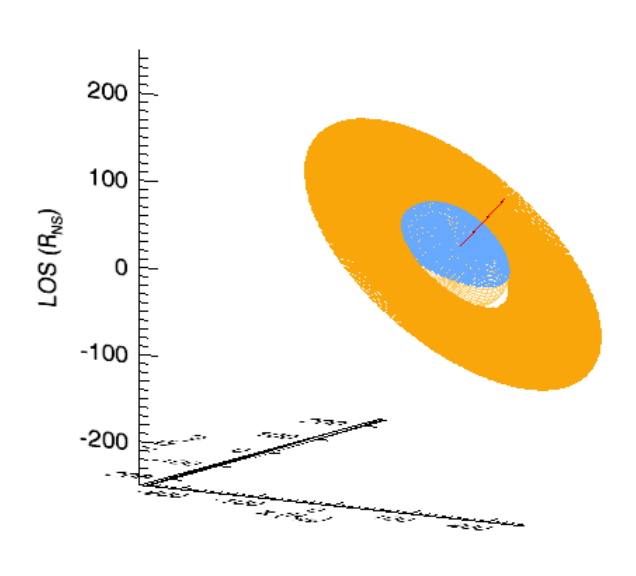}}}
\caption[]{The selection of the part in view for the torus, in panels (a) and (b), while considering also the disk, in panels (c) and (d). In panels (a) and (c) the source is shown exactly along the LOS direction, while in panels (b) and (d) the reference frame is rotated, for ease of visualization.
The points in view are marked in cyan for the torus and in orange for the disk, while those not in view are marked in yellow for the torus and white for the disk. In this case, we used $\chi=45^{\circ}$, $\xi=1^{\circ}$, $R_\mathrm{max}=70\,\rns$, and $R_\mathrm{disk}=200\,\rns$.}
\label{fig:los_pointofview and not}
\end{figure*}

Figure \ref{fig:los_pointofview and not} shows the emitting regions of the source for a given viewing geometry, in the LOS reference frame. 
Although a full treatment requires to take into account general relativistic (GR) effects, such as gravitational redshift and ray-bending, we expect them to be negligible since the emitting regions (torus and disc) lie in general far from the star. For these reasons, we do not consider GR corrections in the present simulations.

\subsubsection{The flux} 
Once a point within the emission region of the torus is tagged as in view, we calculate its temperature. As reported above, the temperature of the torus varies with the magnetic colatitude. The temperature of each emission point is obtained by interpolating over the grid discussed in \cite{brice2023observational}. 
Instead, for a visible point on the disk we derive the temperature from the Shakura-Sunyaev profile (equation \ref{temperatura del disco}), once the inner radius $R_\mathrm{in}$ and the inner temperature $T_\mathrm{in}$ of the disk are fixed (subsection \ref{subsec:spectral analyses}). The locally emitted flux is calculated from the blackbody distribution at the local temperature. We then sum all the contributions coming from different points in view  using the same approach as discussed in \citet[see their section 3.2]{taverna2017spectrum}, obtaining as output a matrix with the total observed flux as a function of energy and phase.

\subsubsection{Polarization observables}
\label{polarization}
For each point in view inside $\rpl$ we set $N_\mathrm{X}$ ($N_\mathrm{O}=1-N_\mathrm{X}$) and compute the rotated Stokes parameters of the photons at the adiabatic radius \cite[equation 5 in][]{taverna2015polarization}. As described above for the total flux, we sum the Stokes parameter fluxes over the part in view of the emitting region (at each energy and phase), obtaining the total $Q$ and $U$ fluxes for both X and O photons. Hence, the code calculates PD and PA following equation (\ref{pol_frac_and_pol_ang}). For the points in view that are at a distance $\gtrsim\rpl$ from the star we do essentially the same. However, since, as described above, outside the adiabatic region the photon electric field direction is frozen with respect to the magnetic field one, we anyway expect to observe a much lower polarization degree for radiation emitted from these points. For this reason, and for the sake of simplicity, for photons coming from outside $\rpl$ we fixed directly $N_\mathrm{X}=0.5$ (considering the $50\%$ of radiation to be X-mode photons, and the other $50\%$ to be O-mode photons, with no privileged polarization mode), which, from equation (\ref{euqzione_pil_intrinseca}) gives $\Pi_\mathrm{L}=0$. 
So, in this work, we analyzed the variation of PA and PD with the energy and phase varying $N_\mathrm{X}$ only for photons coming from the regions of the torus closer to the star surface (and so emitted from inside $\rpl$).

\section{Data reduction}
\label{sec:data}
We analyzed a sample of {\it XMM-Newton} and {\it NuSTAR} observations of M51 ULX-7 and NGC 7793 P13, two PULXs taken as benchmarks for our study. We used three {\it XMM-Newton} observations for M51 ULX-7 (Obs.ID: 0824450901, 0830191501 and 0830191601), carried out between 2018 May and June, and two for NGC 7793 P13 (Obs.ID: 0693760401 and 0748390901), carried out between 2013 November and 2014 December. We extracted the EPIC-pn data using the {\sc SAS} v.14.0. package\footnote{Spectra reduced with this version were already available in our archives. Since they are of enough high quality for the aims of this work and since we do not expect significant differences, we choose not to reprocess the data with the latest SAS v21.0.0.}. Spectra were obtained by selecting events with the {\sc pattern} $\leq 4$ for EPIC-pn (single- and double-pixel events) and setting `{\sc flag}=0' to ignore bad pixels and events coming from the CCD edges. Epochs of high background were also removed from the analysis. Source and background events were extracted from circular regions with radii of $30''$ and $60''$, respectively. The spectra were rebinned to have at least 25 counts per energy bin.

\section{Results}
\label{sec:results}
We started by comparing the light curves of the {\it XMM-Newton} observations with the simulated ones to constrain the geometry of view, which significantly affects the flux modulation. In particular, we compared the observed and simulated pulsed fractions (PFs), defined as
\begin{equation}
   \rm{PF} = \frac{\rm{max(Flux)} - \rm{min(Flux)}}{\rm{max(Flux)} + \rm{min(Flux)}}.
\end{equation}
After constraining the viewing geometry, we compared the spectra determining the temperature at the inner radius of the accretion disc ($T_{\rm{in}}$), the torus temperature range, and the magnetic field strength. For the comparison of the pulse profiles, we considered six possible viewing geometries while, for the comparison of the spectra, we chose 16 values of $T_{\rm{in}}$, selected on the basis of values found in the literature, for each different value of the magnetic field strength. We then analyzed the polarization degree and polarization angle for each source using the best-fit viewing geometry, $T_{\rm{in}}$, and magnetic field strength.

\subsection{Comparison of the light curves}
\label{sec:comparison_lightcurves}
We made a direct comparison of the light curves, considering the 2–3 keV and 3–4 keV energy bands since we expect that at these energies the torusdisk model (in particular the emission of the torus) dominates. We compared each observed profile with six simulations having different geometries of view, each differing only for the angle $\chi$ (ranging from $10^\circ$ to $60^\circ$). The value of $\xi$ was fixed at $10^\circ$, because for values of $20^\circ$, or higher, and for values lower than $10^\circ$, the PFs were not compatible with that observed. So, for the sake of simplicity, we fixed $\xi$ at $10^\circ$, also to represent a general non-aligned magnetic rotator case. $\chi$ angles beyond $60^\circ$ were not considered, because we noted in the simulations that the PF decreases if $\chi>60^\circ$. 

\subsubsection{M51 ULX-7}
\label{lc_M51}
To perform simulations we fixed some input parameters: the magnetic field strength $B$, the Alfvén radius $\ra$, the torus temperature, and the disk radius $R_{\rm{disk}}$. M51 ULX-7 has a surface magnetic field strength between $8 \times 10^{11}$ G and $10^{13}$ G \cite[values derived by][from the \textit{P-$\dot{P}$} relation]{castillo2020discovery}, so we set a lower and upper limit, $B = 10^{12}$ G and $B = 8 \times 10^{12}$ G, respectively. The Alfvén radius ($\ra$, which coincides with the inner radius of the disk $R_{\rm{in}}$) and the temperature of the torus were calculated according to $B$. The outer radius of the disk was set to $200R_{\rm{NS}}$, as beyond this radius the emitted flux falls below 0.1 keV, outside the range considered here. We report all the input fixed parameters in table \ref{tabella_temp_rmax1}.

\begin{figure*}
\centering
    \includegraphics[scale=0.5]{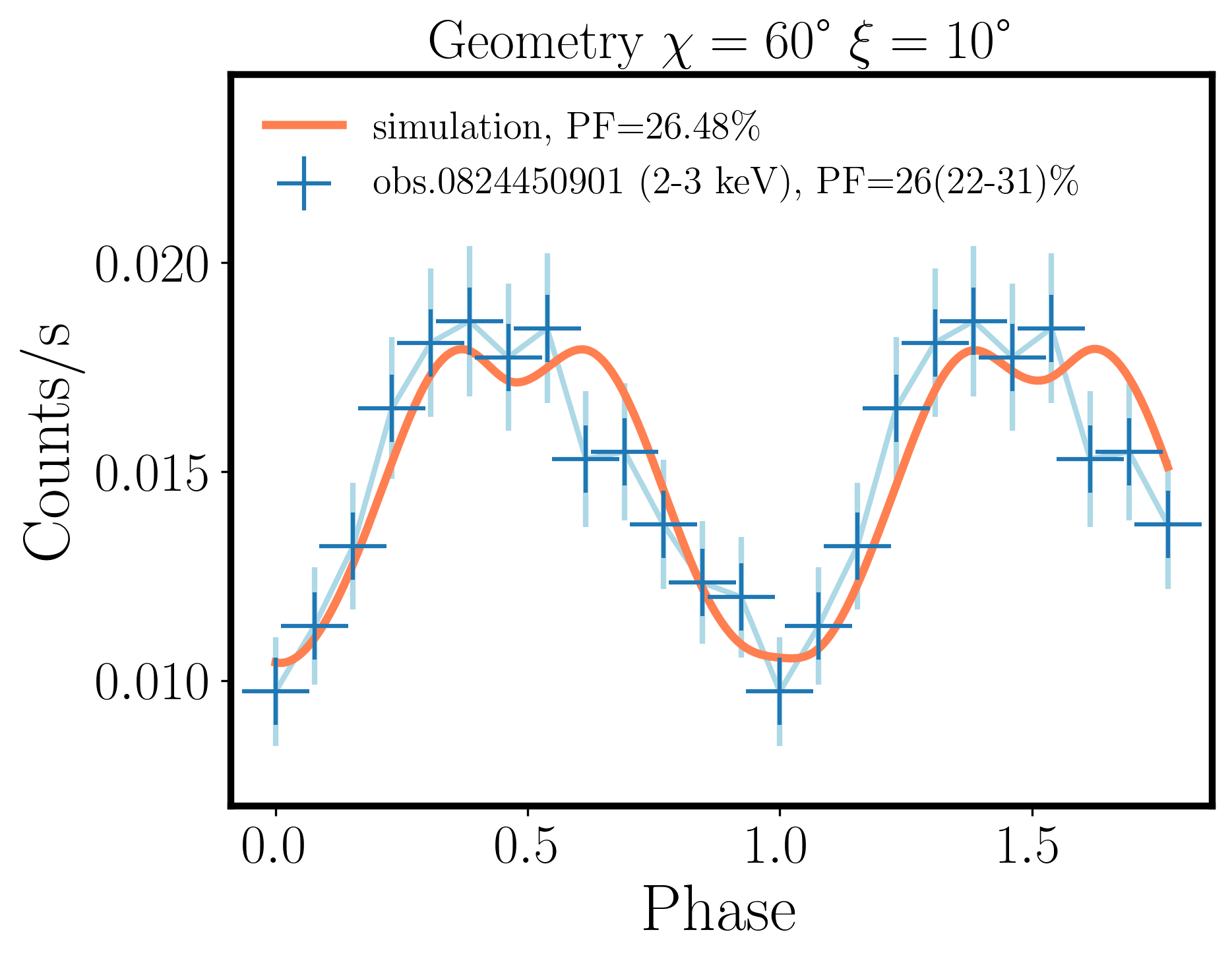}
    \includegraphics[scale=0.5]{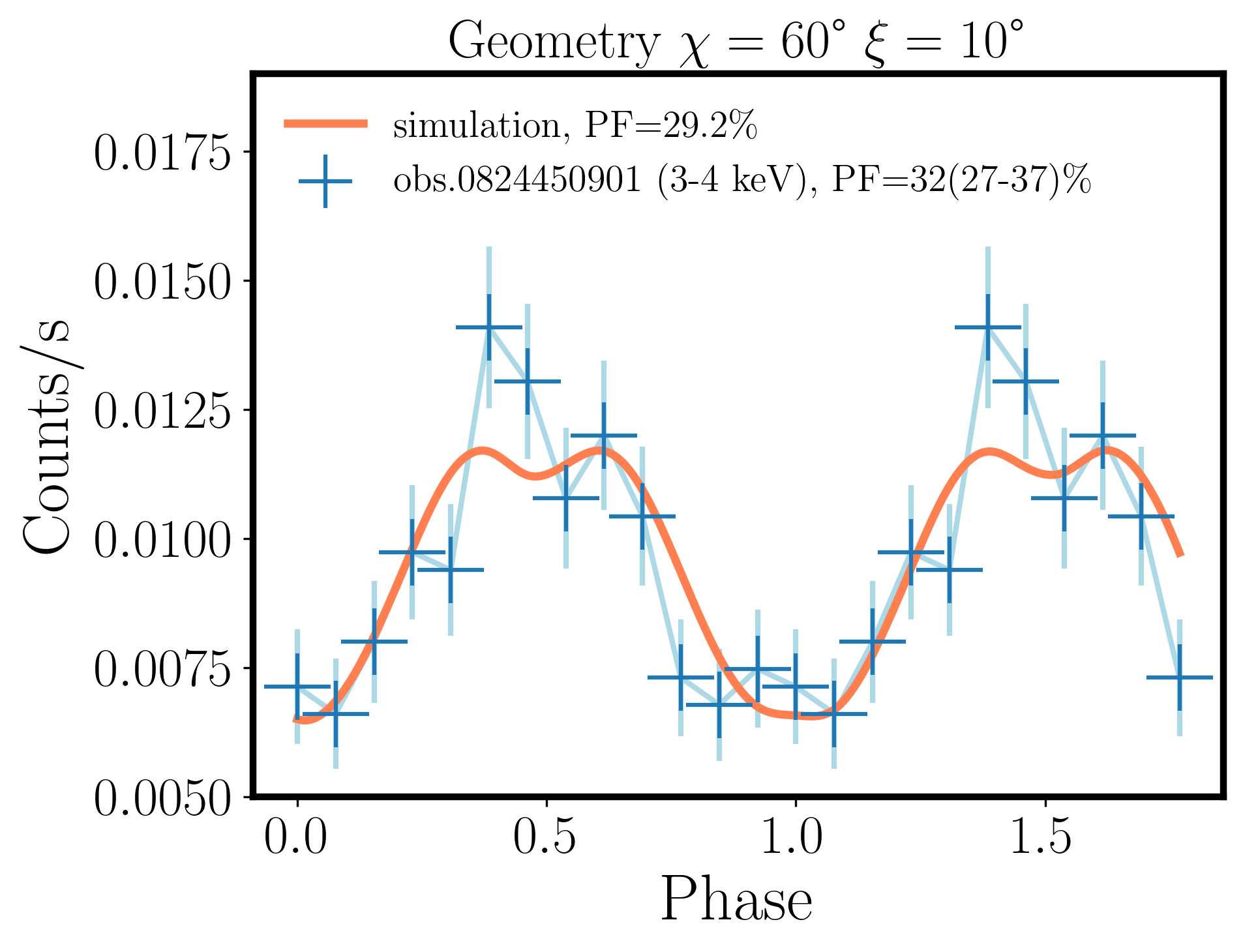}\\
    \includegraphics[scale=0.5]{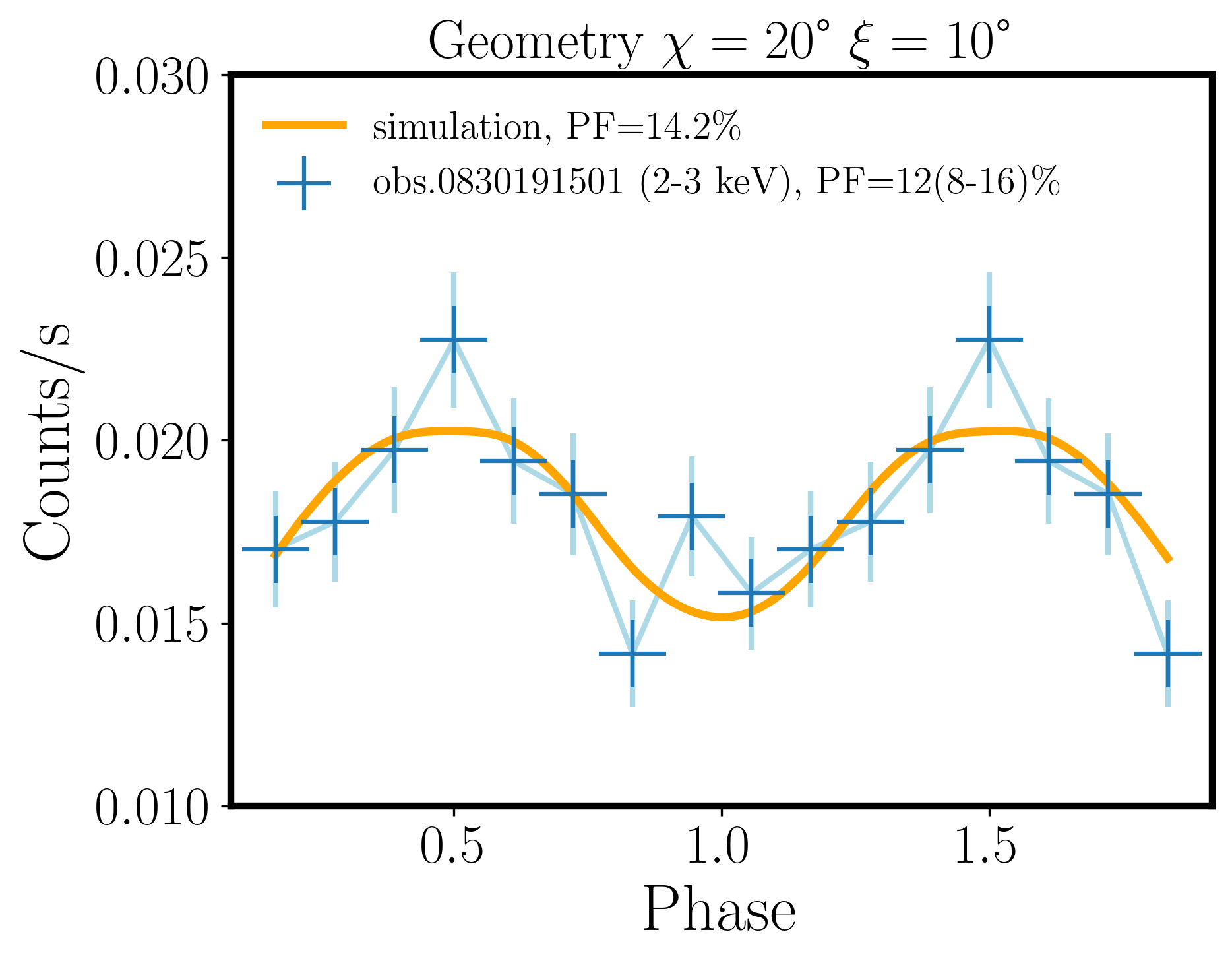}
    \includegraphics[scale=0.5]{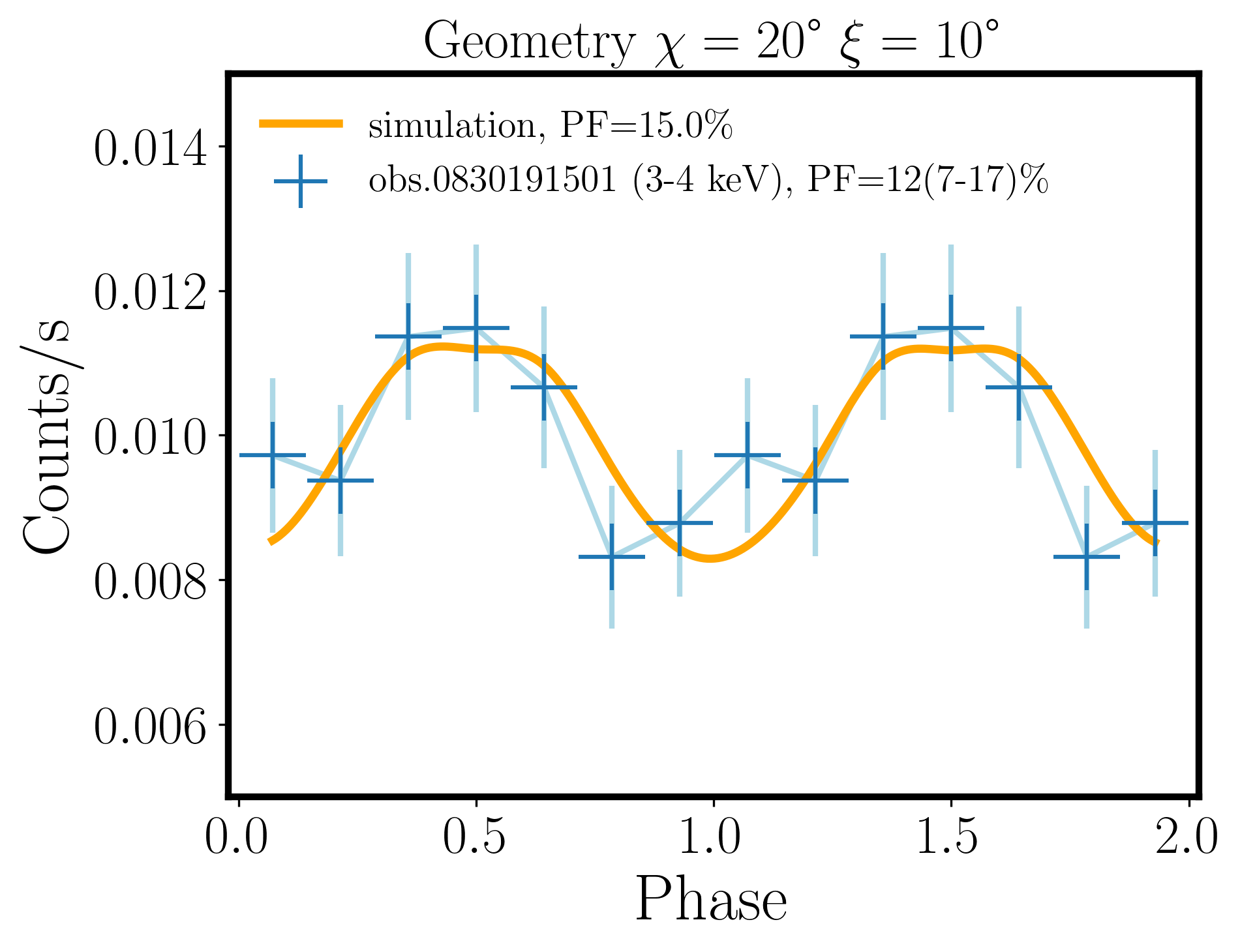}\\
    \includegraphics[scale=0.5]{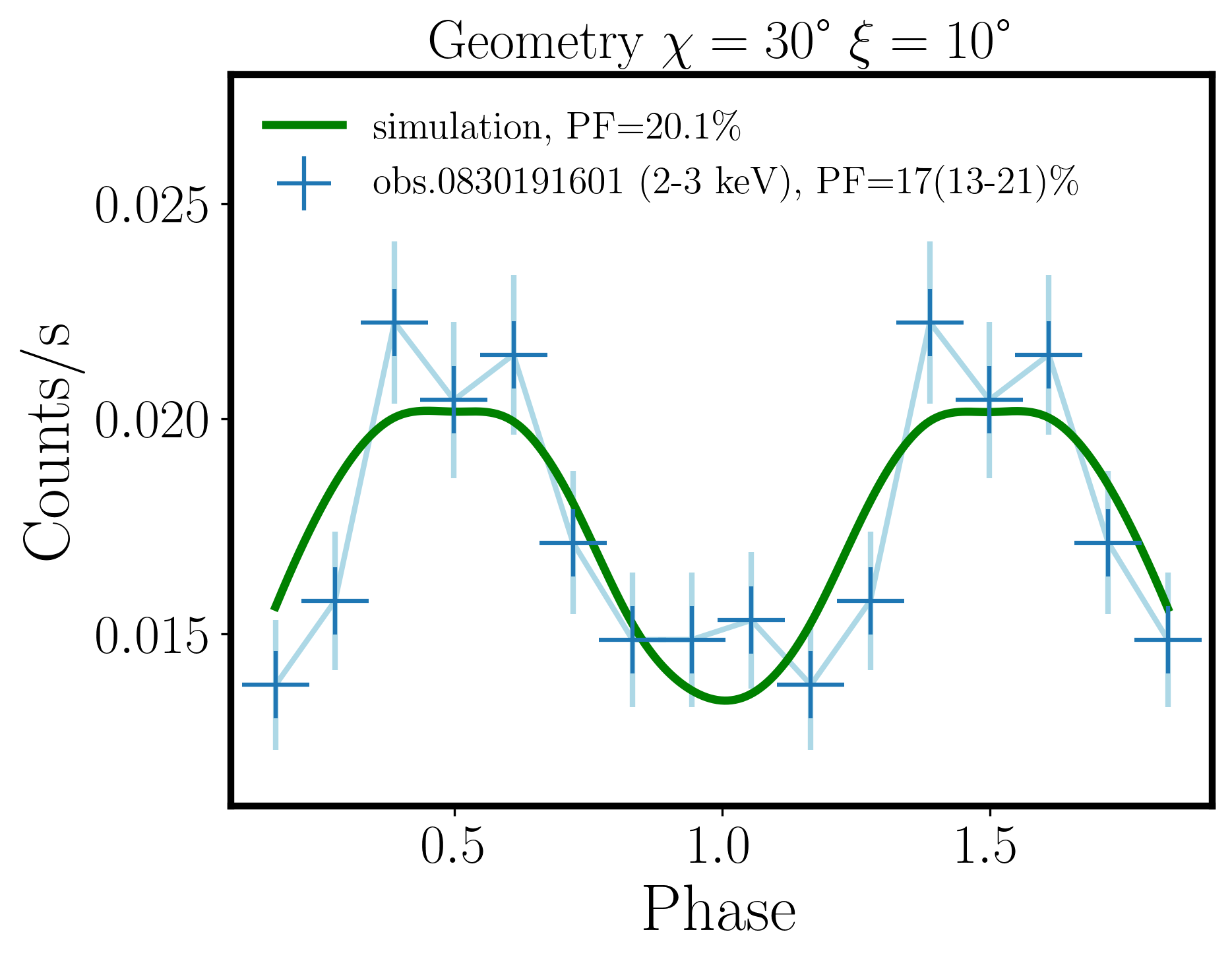}
    \includegraphics[scale=0.5]{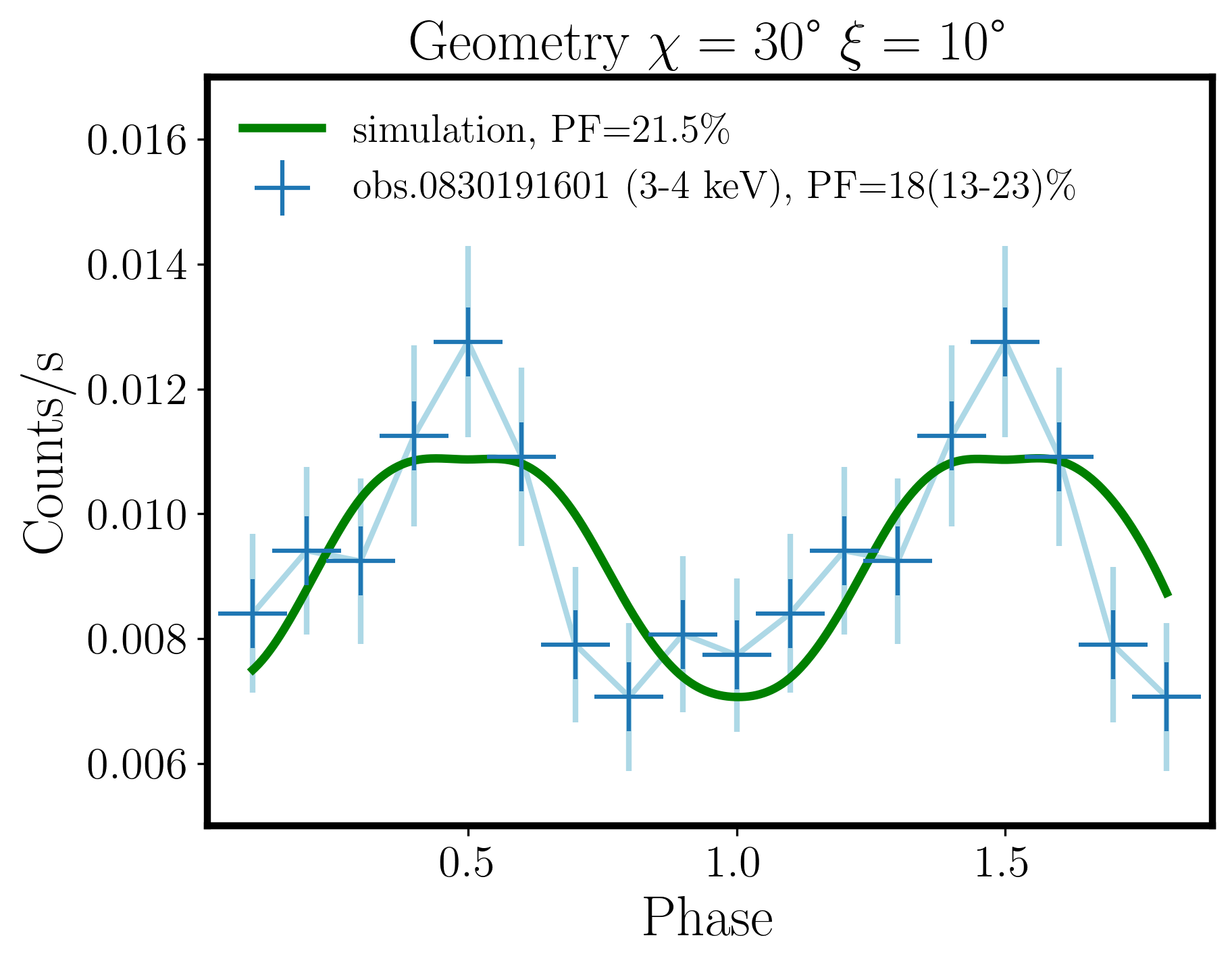}
    \caption{Comparison of the three {\it XMM-Newton} light curves of M51 ULX-7 with simulations. From top to bottom geometry of view with $\chi=60^\circ$ $\xi=10^{\circ}$ (coral), with $\chi=20^\circ$ $\xi=10^{\circ}$ (orange), and with $\chi=30^\circ$ $\xi=10^{\circ}$ (green). The column on the left refers to the energy band 2--3 keV, while that on the right to the energy band 3--4 keV.}
\label{figure_confronti_pf}
\end{figure*}

For M51 ULX-7, we found a lower and upper limit for the geometry of view, $\chi=20^\circ$ $\xi=10^\circ$ and $\chi=60^\circ$ $\xi=10^\circ$, respectively (see figure \ref{figure_confronti_pf}). In table \ref{tabella_pf_M51} we report the values of the observed PFs with those of the best-fit viewing geometries, across the two different energy bands. The observed PFs show general agreement with the simulated ones within uncertainties.
\begin{table}[!ht]
    \centering
    \renewcommand\arraystretch{1.6}
    \begin{tabular}{c c|c|c}
    \hline
    \multicolumn{1}{c}{$B$ (G)} & \multicolumn{1}{c|}{$T_\mathrm{torus,max}$ (keV)} & \multicolumn{1}{c}{$T_\mathrm{torus,min}$ (keV)} & \multicolumn{1}{|c}{$\ra$ ($R_\mathrm{NS}$)} \\  
    \hline
    $10^{12}$ & 1.38 & 1.06 & 19 \\
    $8\times10^{12}$ & 0.77 & 0.57 & 52\\
    \hline
    \end{tabular}
    \caption{Magnetic filed strength $B$, minimum and maximum temperature of the torus, $T_\mathrm{torus,min}$ and $T_\mathrm{torus,max}$ respectively, and the Alfvén radius, $\ra$, for spectral simulations of M51 ULX-7.}
    \label{tabella_temp_rmax1}
\end{table}
\begin{table*}[!ht]
    \centering
    \renewcommand\arraystretch{1.6}
    \begin{tabular}{c c|c|c|c}
    \hline
    \multicolumn{1}{c}{Obs. ID} & \multicolumn{2}{c|}{Observed PFs ($\%$)} & \multicolumn{2}{c}{Simulated PFs ($\%$)} \\  
     & \multicolumn{1}{c}{2--3 keV} & \multicolumn{1}{c}{3--4 keV} & \multicolumn{1}{|c}{2--3 keV} & \multicolumn{1}{c}{3--4 keV}\\
    \hline
      0824450901 & $26^{31}_{22}$  & $32^{27}_{37}$ & 26.48\tablefootmark{d} & 29.2\tablefootmark{d} \\
      0830191501 & $12^{16}_{8}$ & $12^{17}_{7}$  & 14.2\tablefootmark{a}  & 15.0\tablefootmark{a}  \\
      0830191601 & $17^{21}_{13}$ & $18^{23}_{13}$ & 20.1\tablefootmark{b}  & 21.5\tablefootmark{b} \\
    \hline
    \end{tabular}
    \caption{Observed pulsed fractions of M51 ULX-7 for the three \textit{XMM-Newton} observations, against simulated ones both calculated in the energy bands 2--3 keV and 3--4 keV.}
    \label{tabella_pf_M51}
\tablefoot{\\
\tablefoottext{a}{Geometry of view with $\chi=20^\circ$}\\
\tablefoottext{b}{Geometry of view with $\chi=30^\circ$}\\
\tablefoottext{c}{Geometry of view with $\chi=60^\circ$}}
\end{table*}
In figure \ref{figure_confronti_pf} we show the comparisons.

\subsubsection{NGC 7793 P13}
\label{lc_P13}
\begin{figure*}
\centering
    \includegraphics[scale=0.5]{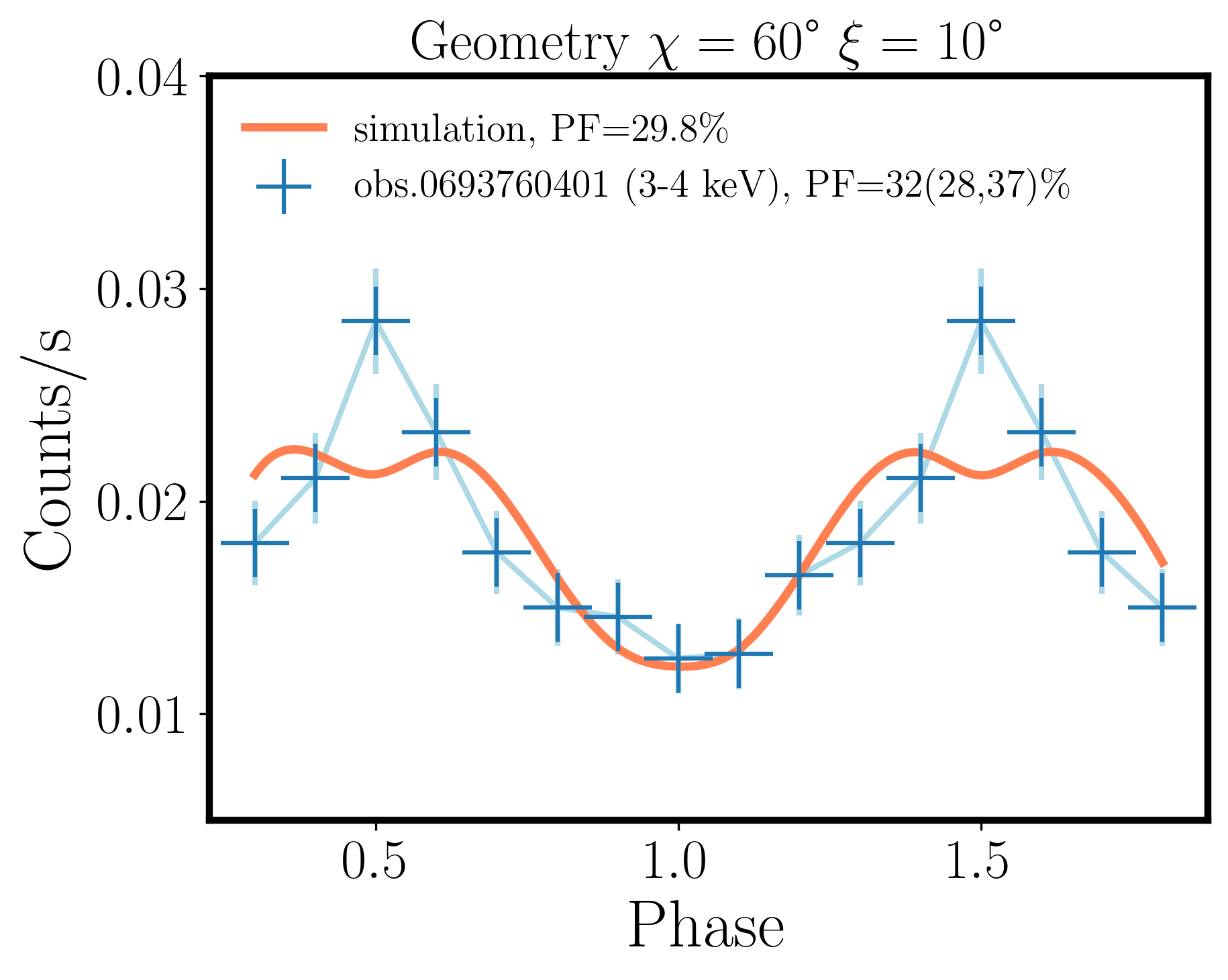}
    \includegraphics[scale=0.5]{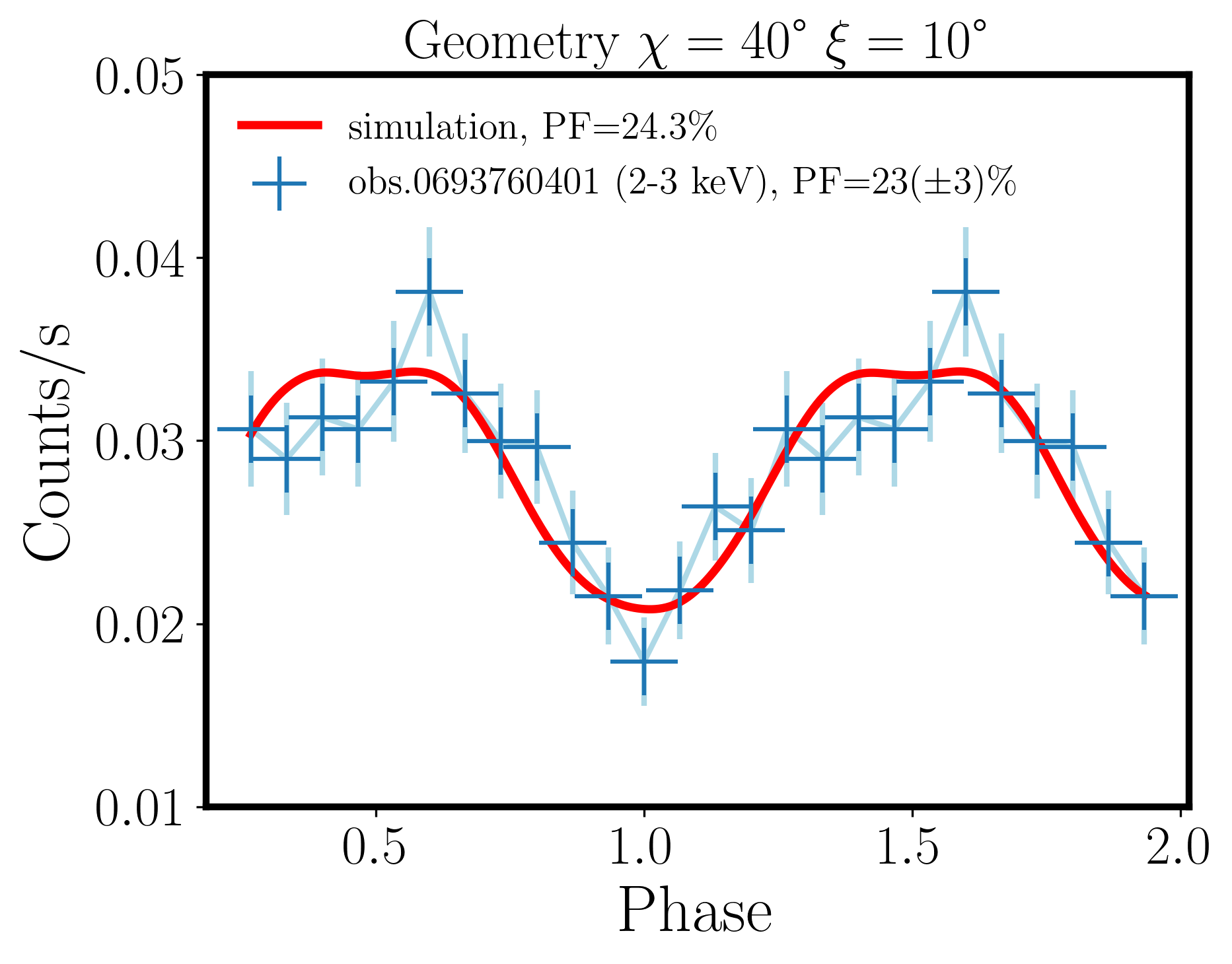}
    \includegraphics[scale=0.5]{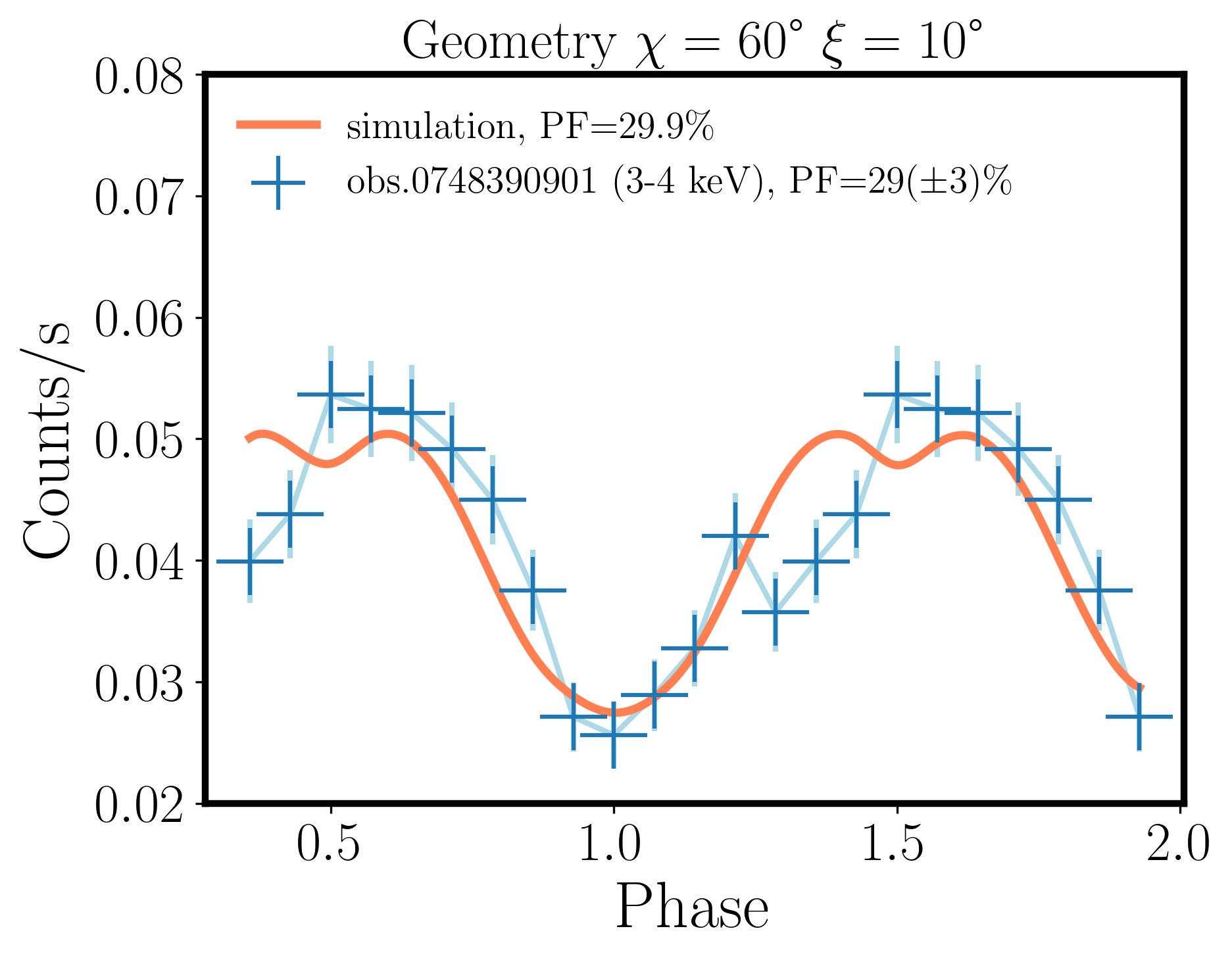}
    \includegraphics[scale=0.5]{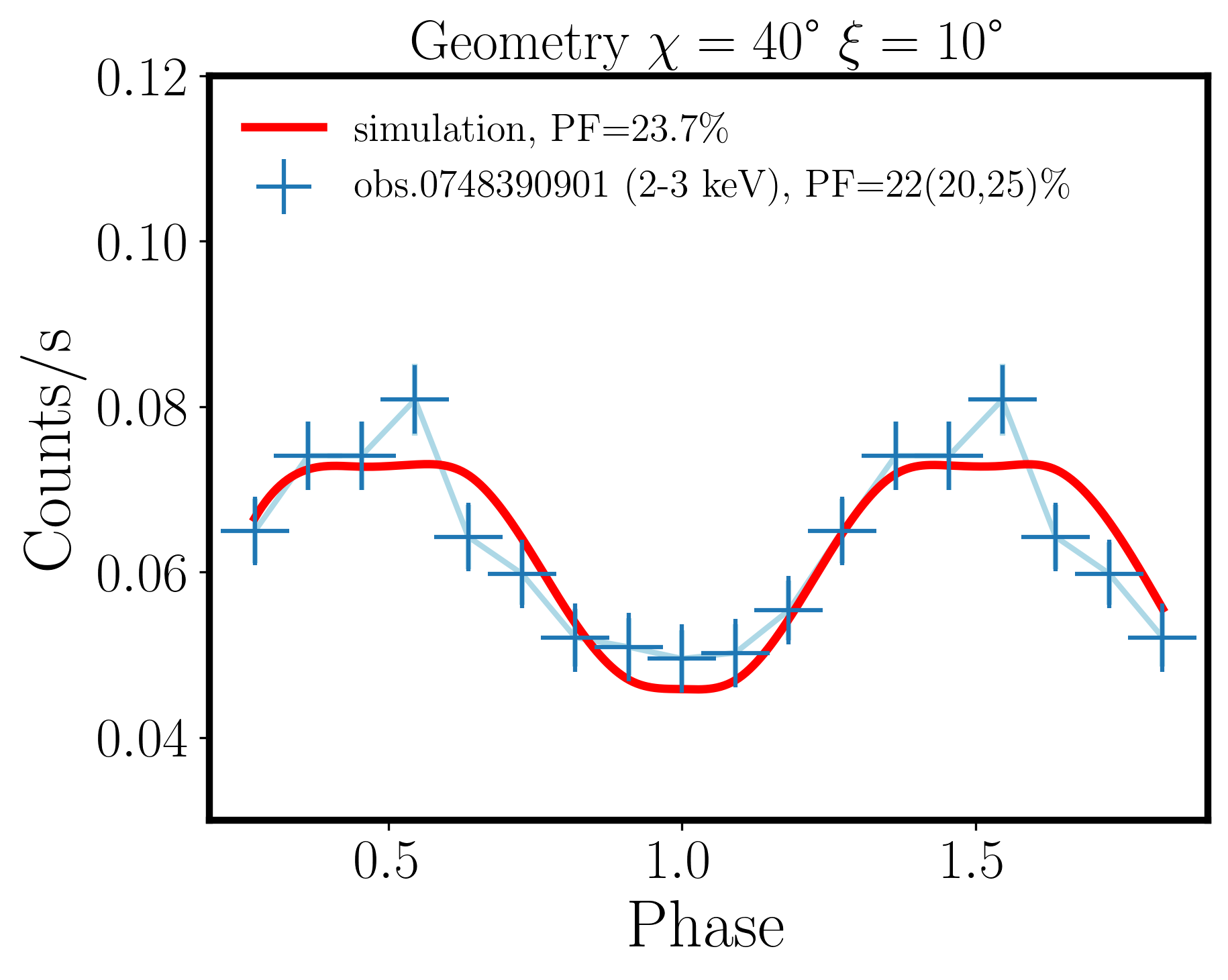}
    \caption{Comparison of the two {\it XMM-Newton} light curves of NGC 7793 P13 with simulations. From left to right the geometry of view with $\chi=60^\circ$ $\xi=10^{\circ}$ (coral) and with $\chi=40^\circ$ $\xi=10^{\circ}$ (red). The column on the right refers to the energy band 2--3 keV, while that on the left refers to the energy band 3--4 keV.}
\label{figure_confronti_pf_P13}
\end{figure*}

We did the same for NGC 7793 P13, considering that the observed surface magnetic dipole field strength is around $2\times 10^{12}$ G \cite[][still obtained from $P$ and $\dot{P}$]{furst2016discovery,israel2017discovery}. We then set the dipole magnetic field strength to $B=4\times10^{12}\,\rm{G}$ considering the value at the poles, and computed the Alfvén radius of the torus, the torus temperatures (see table \ref{tabella_temp_P13}), and set the disk outer radius to $R_\mathrm{disk}=200\,\rns$. 
\begin{table}[!ht]
    \centering
    \renewcommand\arraystretch{1.6}
    \begin{tabular}{c c|c|c}
    \hline
    \multicolumn{1}{c}{$B$ (G)} & \multicolumn{1}{c|}{$T_\mathrm{torus,max}$ (keV)} & \multicolumn{1}{c}{$T_\mathrm{torus,min}$ (keV)} & \multicolumn{1}{|c}{$\ra$ ($R_\mathrm{NS}$)} \\  
    \hline
    $4\times10^{12}$ & 0.79 & 0.19 & 49 \\
    \hline
    \end{tabular}
    \caption{Magnetic filed strength $B$, minimum and maximum temperature of the torus, $T_\mathrm{torus,min}$ and $T_\mathrm{torus,max}$, respectively, and the Alfvén radius, $\ra$, for spectral simulations of NGC 7793 P13.}
    \label{tabella_temp_P13}
\end{table}
This time we found much less variation for the angle $\chi$, identifying $\chi=60^\circ$ $\xi=10^{\circ}$ and $\chi=40^\circ$ $\xi=10^{\circ}$ as the upper and lower limit, respectively. In figure \ref{figure_confronti_pf_P13} the light curves and in table \ref{tabella_pf_P13} the pulsed fractions, across the two energy bands, are confronted with the observed ones.
\begin{table*}[!ht]
    \centering
    \renewcommand\arraystretch{1.6}
    \begin{tabular}{c c|c|c|c}
    \hline
    \multicolumn{1}{c}{Obs. ID} & \multicolumn{2}{c|}{Observed PFs ($\%$)} & \multicolumn{2}{c}{Simulated PFs ($\%$)} \\  
     & \multicolumn{1}{c}{2--3 (keV)} & \multicolumn{1}{c}{3--4 (keV)} & \multicolumn{1}{|c}{2--3 (keV)} & \multicolumn{1}{c}{3--4 (keV)}\\
    \hline
      0693760401 & $23^{26}_{19}$ & $32^{37}_{28}$  & 24.3\tablefootmark{a} & 29.8\tablefootmark{b} \\
      0748390901 & $22^{25}_{20}$ & $29^{32}_{26}$  & 23.7\tablefootmark{a} & 29.9\tablefootmark{b} \\
    \hline
    \end{tabular}
    \caption{Observed pulsed fractions of NGC 7793 P13 for the two \textit{XMM-Newton} observations against simulated ones both calculated in the energy bands 2--3 keV and 3--4 keV.}
    \label{tabella_pf_P13}
\tablefoot{\\
\tablefoottext{a}{Geometry of view with $\chi=40^\circ$}\\
\tablefoottext{b}{Geometry of view with $\chi=60^\circ$}}
\end{table*}

\subsection{Comparison of the spectra}
\label{subsec:spectral analyses}
\subsubsection{M51 ULX-7}
In section \ref{lc_M51} we derived a lower and upper limit for the geometry of view of the source, $\chi=20^\circ$, $\xi=10^\circ$ and $\chi=60^\circ$, $\xi=10^\circ$, respectively. We now perform spectral comparisons to constrain $T_{\rm{in}}$ for both the geometries and the magnetic field strengths. From the literature, the temperature of the soft spectral component (which is related to $T_\mathrm{in}$) varies between 0.33 and 0.5 keV \cite[][]{castillo2020discovery,brightman2022evolution,vasilopoulos2020m51}. We then produced 16 spectral simulations, each with a different temperature value $T_\mathrm{in}$, which ranges between 0.25 and 0.5 keV.

\subsubsection{M51 ULX-7 -- $B=10^{12}$ G}
\begin{figure*}
\centering
    \includegraphics[scale=0.5]{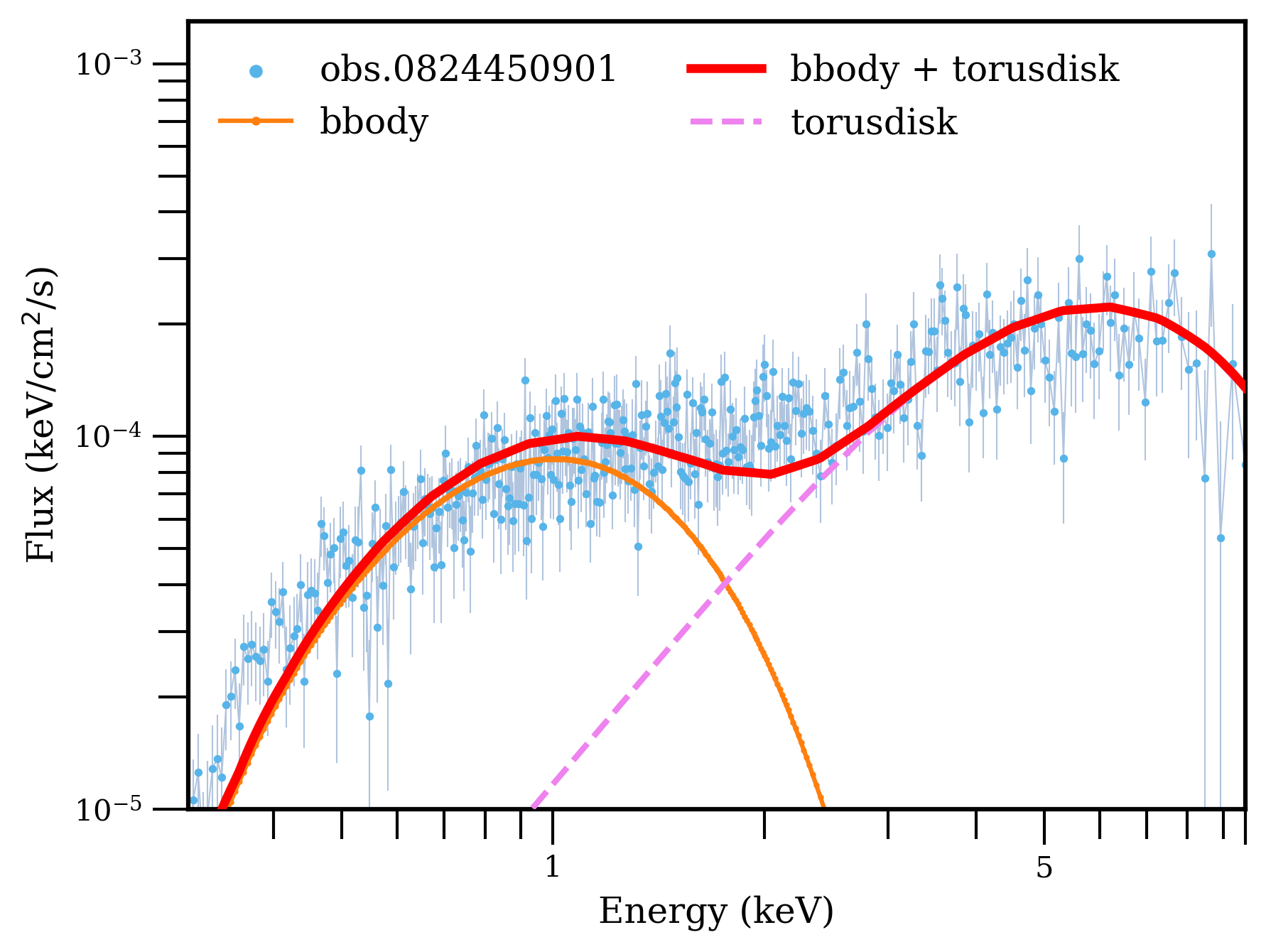}
    \includegraphics[scale=0.5]{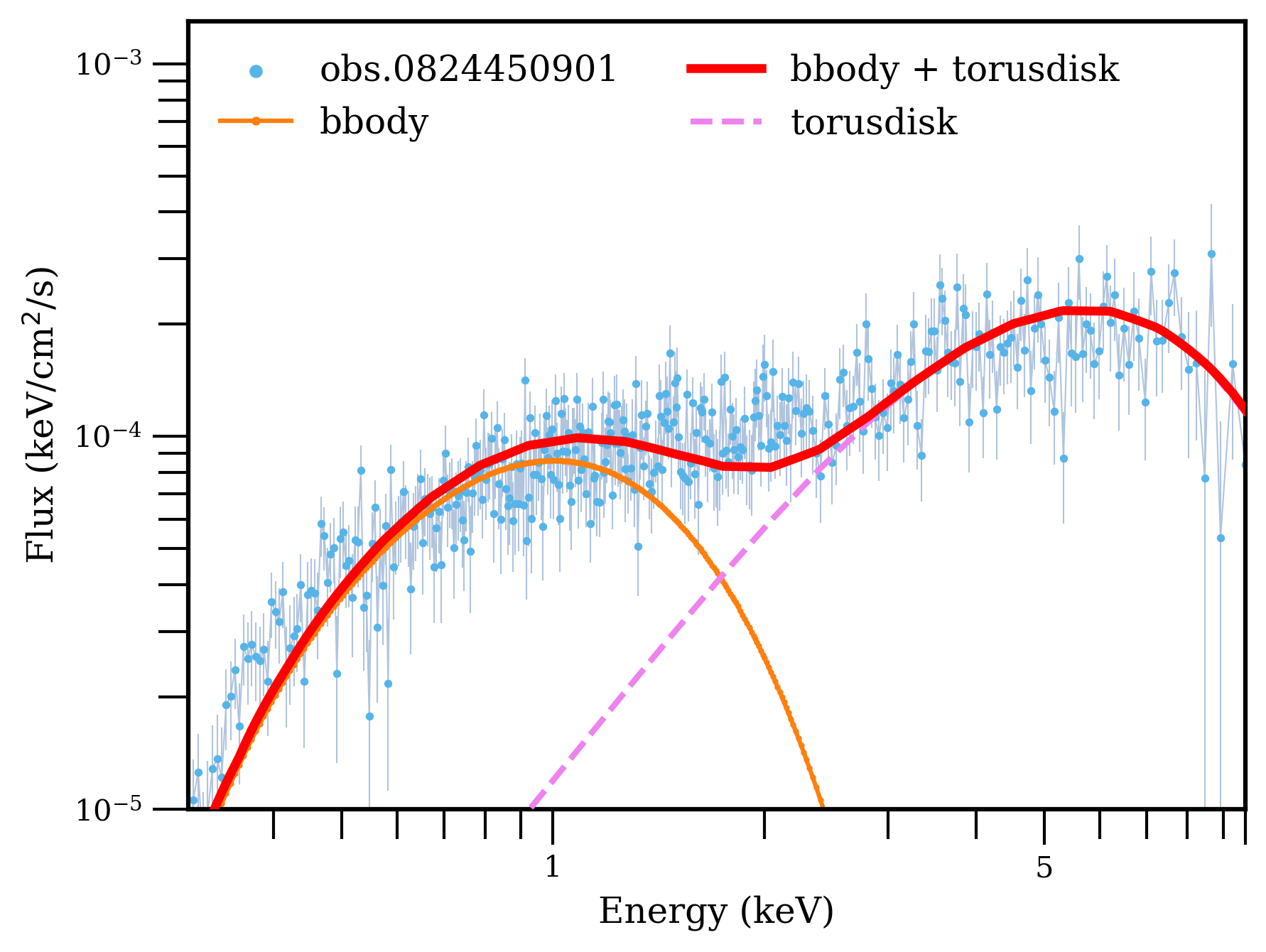}
    \caption{X-ray spectrum of M51 ULX-7 obtained with {\it XMM-Newton} (observation: 0824450901). The spectrum is fitted with a multicomponent model. The panel on the left refers to the viewing geometry $\chi=20^{\circ}$ $\xi=10^{\circ}$, that on the right to $\chi=60^{\circ}$ $\xi=10^{\circ}$. One component is phenomenological ({\sc bbody}), and one is calculated with the model {(\sc torusdisk)}.}
\label{M51_fit_B1}
\end{figure*}

We reproduced the spectrum with the superposition of two thermal components, the XSPEC {\sc bbody} model \cite[see][]{arnaud1996astronomical} and our simulated multicolor blackbody, similarly as in \cite{castillo2020discovery}. We considered the interstellar absorption through the {\sc wabs} model \cite[][]{morrison1983interstellar} leaving the column density as a free parameter of the fit. We carried out the fitting procedure outside XSPEC, using a Python script where we implemented the analytical expressions of {\sc bbody}. This is done to include the {\sc torusdisk} contribution (i.e. the output of the ray-tracer code). For the same reason, we use {\sc wabs} and not {\sc tbabs}, as an analytical expression is available for the former. We fixed the emission temperature of the {\sc bbody} at $0.25$ keV, to ensure the convergence of the fit\footnote{The value of the {\sc bbody} temperature is calculated by separately fitting this component in the energy range where it dominates.}.
Furthermore, we stress that our purpose is to test the reliability of the {\sc torusdisk} model by varying the contribution of the internal disk temperature $T_\mathrm{in}$ for a given viewing geometry, while accounting for the presence of additional components. 
In figure \ref{M51_fit_B1} we show the spectral fits for both the geometries of view of the first observation (the others provide similar results). In table \ref{table_M51_B1} we report the best-fit parameters of all the observations. The best value for the temperature $T_\mathrm{in}$ turns out to be 0.5 keV, in agreement with \cite{castillo2020discovery} within the uncertainties. Despite the superposition of these two thermal components can fit the data, it also leaves non negligible residuals between $1.5$--$3$ keV.  

\begin{table*}
\caption{Best fit parameters for M51 ULX-7 ($B=10^{12}$ G) }
\label{table_M51_B1}      
\centering                          
\renewcommand\arraystretch{1.2}
\begin{tabular}{ c c c c c c }     
 \hline \hline
& \multicolumn{5}{c}{$\chi=20^\circ$, $\xi=10^\circ$} \\
\multicolumn{1}{c}{Observation ID} & N\tablefootmark{a} ($10^{23}$) & J\tablefootmark{b} ($10^{-3}$) & $T_\mathrm{in}$\tablefootmark{c} & $N_\mathrm{H}$ &  $\chi^{2}$/dof \\
 & ($\rm{keV/cm^2/s}$) & ($\rm{keV/cm^2/s}$) &  (keV) &($10^{22} \mathrm{cm^{-2}}$) & \\
 \cmidrule(r){2-6} \\
 0824450901 & 4.17 $\pm$ 0.07 & 4.8 $\pm$ 0.2 & 0.5   & 3.3 $\pm$ 1.4    & 629/386  \\ 
 0830191501 & 3.17 $\pm$ 0.06 & 6.6 $\pm$ 0.3 & 0.5   & 4.0 $\pm$ 1.6    & 647/382  \\ 
 0830191601 & 3.27 $\pm$ 0.06 & 4.5 $\pm$ 0.2 & 0.5   & 2.4 $\pm$ 1.4    & 619/382  \\
 \\
 & \multicolumn{5}{c}{$\chi=60^\circ$, $\xi=10^\circ$} \\
\multicolumn{1}{c}{Observation ID} & N\tablefootmark{a} ($10^{23}$) & J\tablefootmark{b} ($10^{-3}$) & $T_\mathrm{in}$\tablefootmark{c} & $N_\mathrm{H}$  & $\chi^{2}$/dof \\
 & ($\rm{keV/cm^2/s}$) & ($\rm{keV/cm^2/s}$) &  (keV) &($10^{22} \mathrm{cm^{-2}}$)  &  \\
 \cmidrule(r){2-6} \\
 0824450901 & 5.73$\pm$0.09    & 4.7 $\pm$ 0.2   &  0.5  & 2.8 $\pm$ 1.4   &  558/386 \\ 
 0830191501 & 4.37 $\pm$ 0.08  & 6.5 $\pm$ 0.3   &  0.5  & 3.5 $\pm$ 1.5   &  584/382 \\ 
 0830191601 & 4.51 $\pm$ 0.07  & 4.4 $\pm$ 0.2   &  0.5  & 1.8 $\pm$ 1.3   &  555/382 \\
 \hline
\end{tabular}
\tablefoot{\\
\tablefoottext{a}{{\sc torusdisk} norm}\\
\tablefoottext{b}{  {\sc bbody} norm}\\
\tablefoottext{c}{Temperature at the inner radius of the accretion disk}}
\end{table*}

\subsubsection{M51 ULX-7 -- $B=8\times10^{12}$ G}
In this case, we reproduced the spectrum by the superposition of a soft X-ray thermal component plus a hard X-ray non-thermal one. We fitted the former with the sum of two thermal spectral components, the XSPEC {\sc bbody} model and the {\sc torusdisk} multicolor blackbody, while the latter with a cut-off power law ({\sc cutoffpl} in XSPEC), accounting for the interstellar absorption through the {\sc wabs} model. We carried out the fitting procedure, as before, outside XSPEC, implementing, in this case, also the analytical expressions of {\sc cutoffpl}. Still to ensure the convergence of the fit, we fixed the temperature of the {\sc bbody} and also the column density, $N_\mathrm{H}=9\times10^{19} \,\, \rm{cm^{-2}}$ \footnote{To obtain this value of the column density, $N_{H}$, we fitted singularly the three spectral components, each in the energy range where they dominate. Then we calculated $N_{H}$ as the mean of the three values obtained from the fits.}.

\begin{figure*}
\centering
    \includegraphics[scale=0.5]{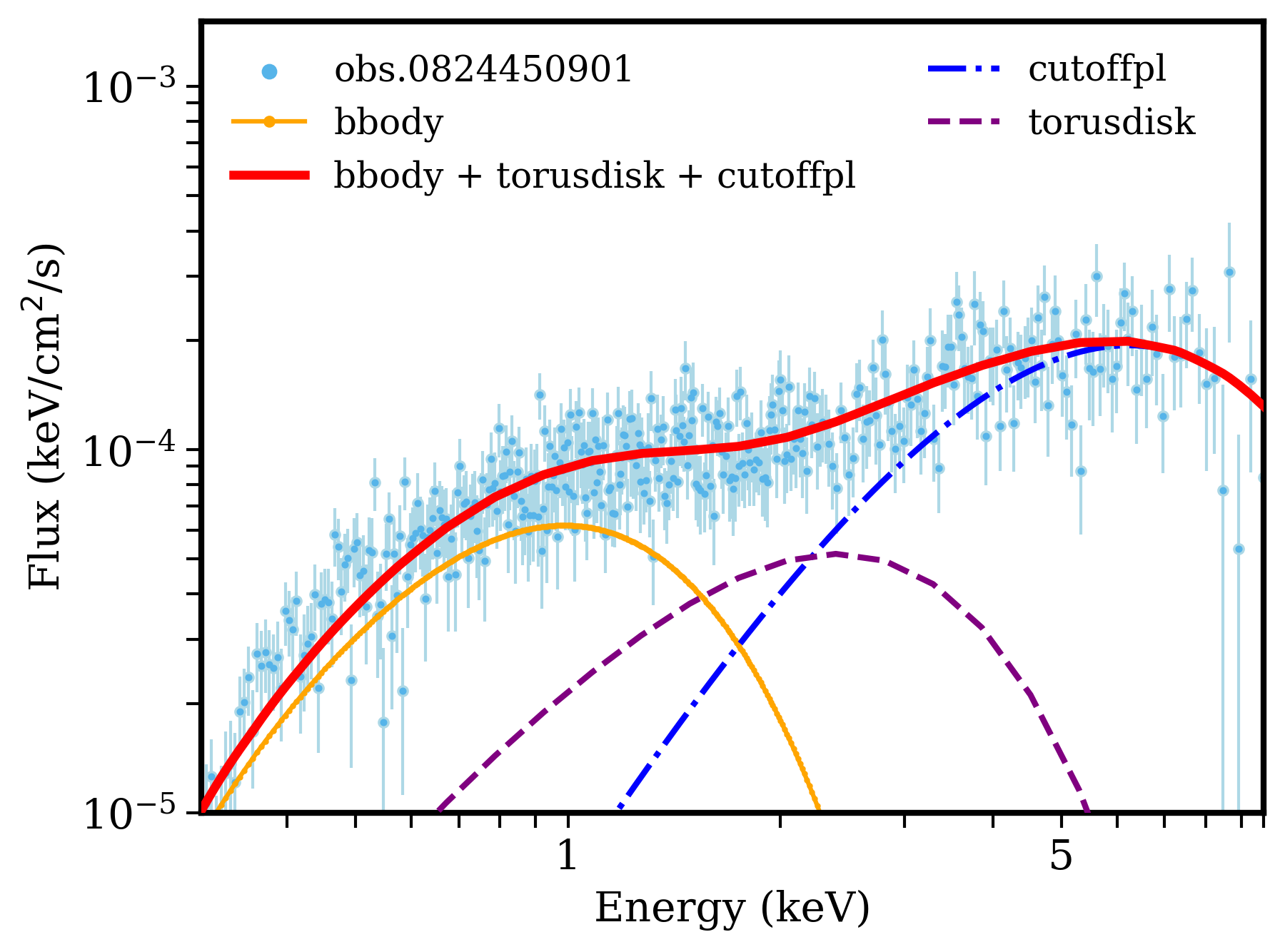}
    \includegraphics[scale=0.5]{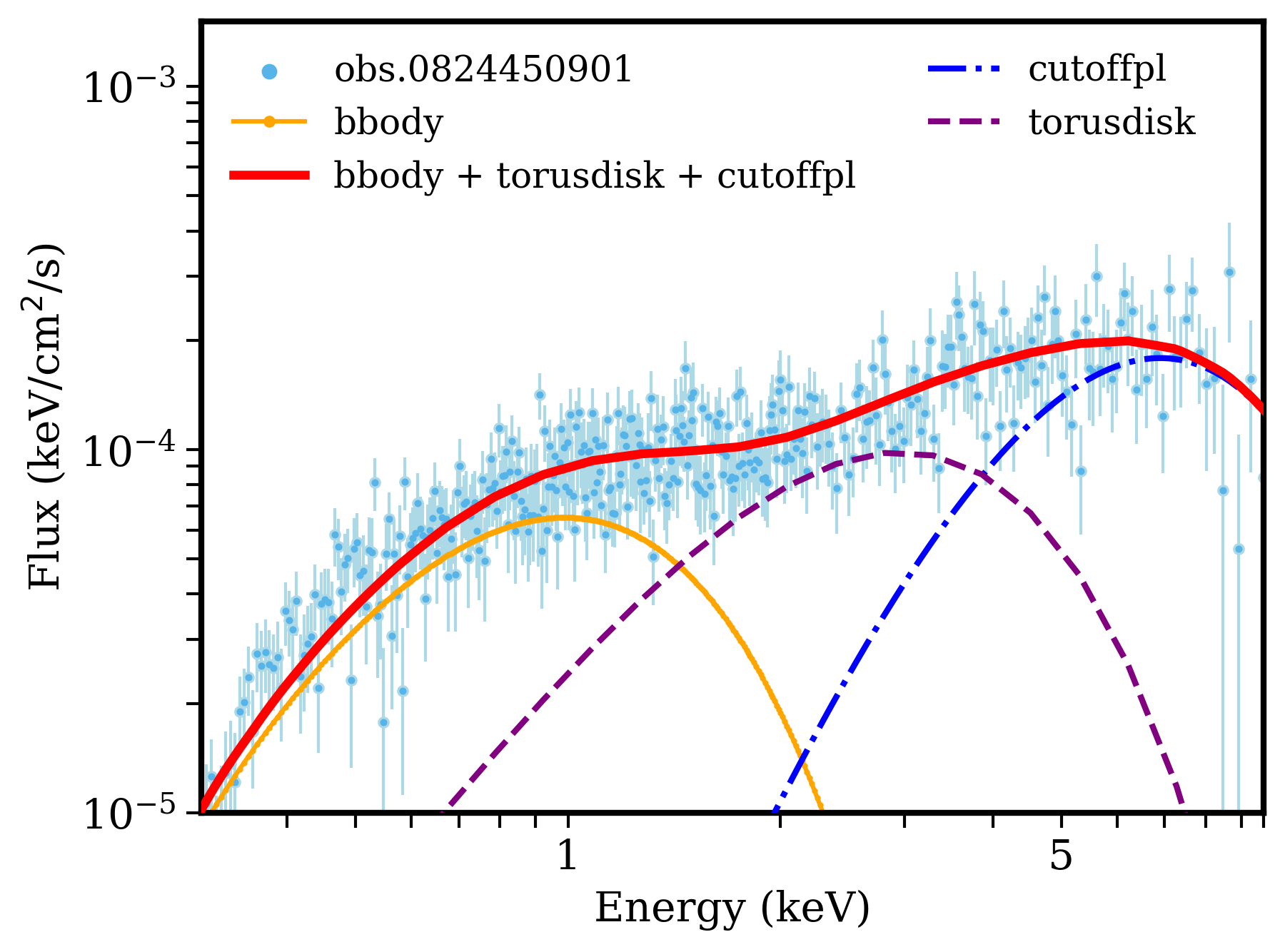}
    \includegraphics[scale=0.5]{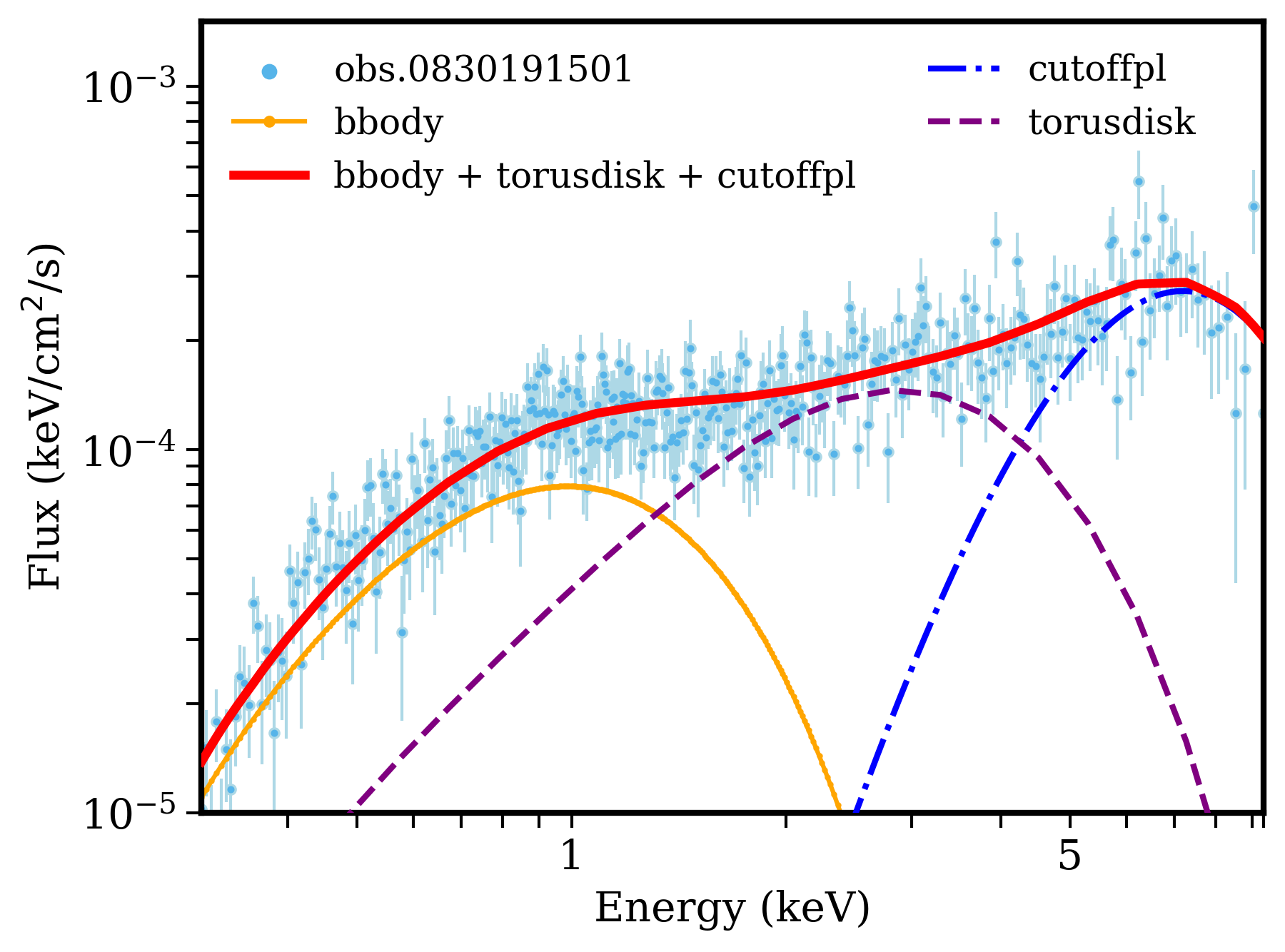}
    \includegraphics[scale=0.5]{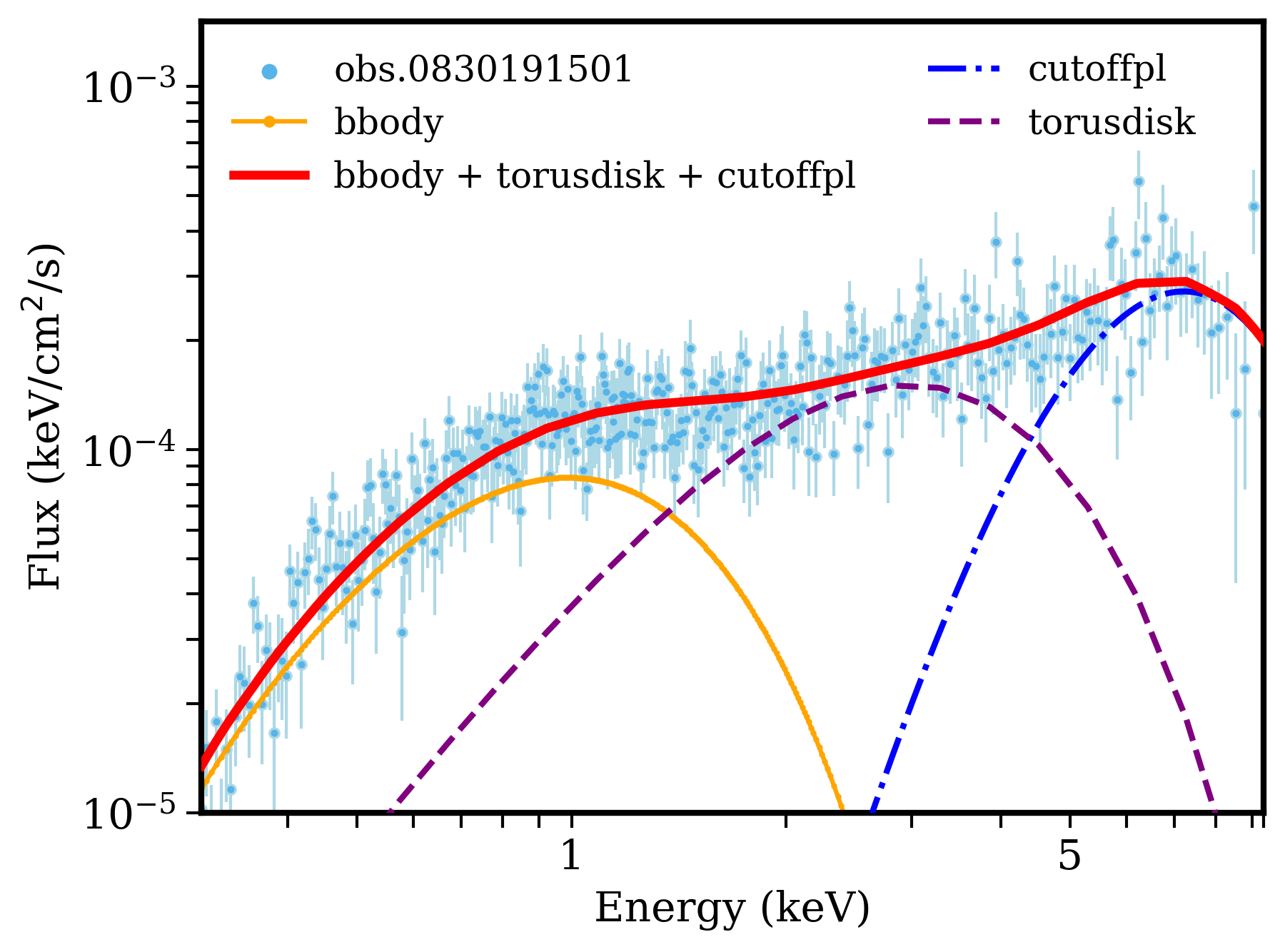}
    \includegraphics[scale=0.5]{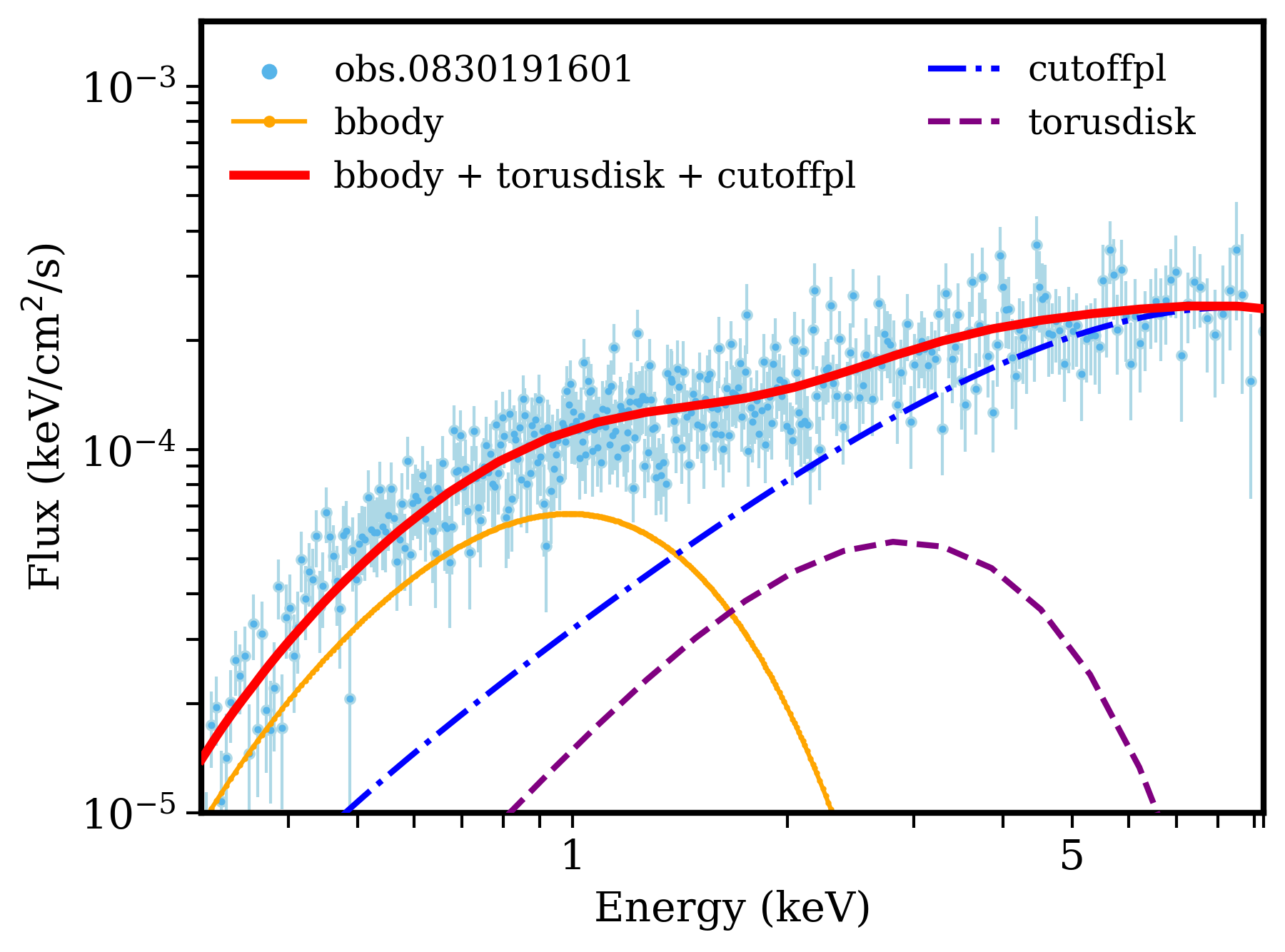}
    \includegraphics[scale=0.5]{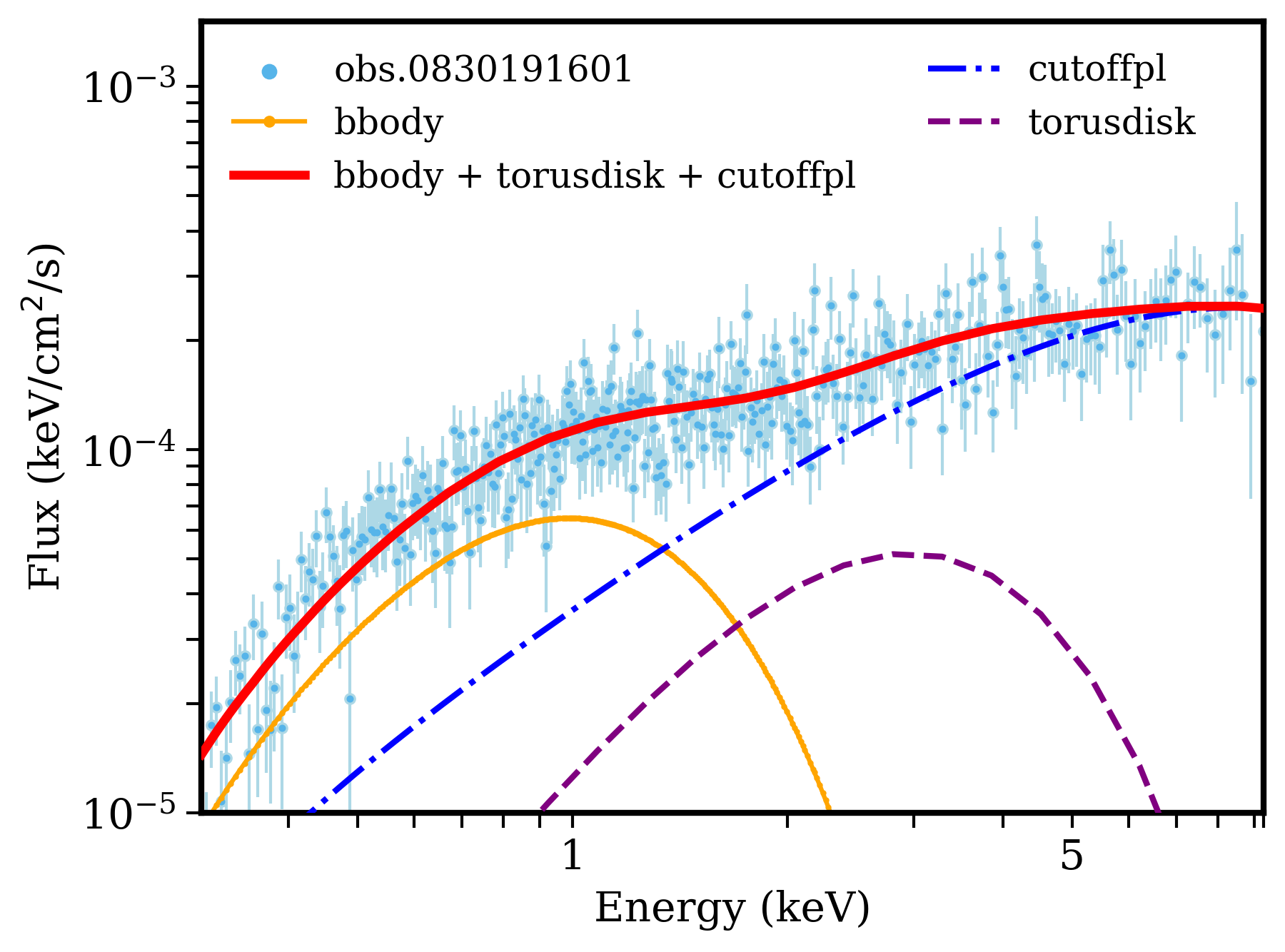}
    \caption{X-ray spectra of M51 ULX-7 obtained with {\it XMM-Newton} (observations: 0824450901, 0830191501, 0830191601). The spectrum is fitted with a multicomponent model. The panels on the left refer to the viewing geometry $\chi=20^{\circ}$ $\xi=10^{\circ}$, while those on the right to $\chi=60^{\circ}$ $\xi=10^{\circ}$. Two components are phenomenological ({\sc bbody}, {\sc cutoffpl}), and one is computed with the model {(\sc torusdisk)}.}
    \label{comparison_spectra_M51}
\end{figure*}

We used the superposition of a blackbody, a multicolor blackbody and a power law component as done by \cite{brightman2022evolution}. The low-energy blackbody component, in fact, is used to describe the emission due to radiative winds that can originate from the accretion disk at super-Eddington accretion rates, \cite[see e.g.][]{poutanen2007supercritically,pinto2016resolved,pinto2020thermal}. In figure \ref{comparison_spectra_M51} we show the spectra for the three {\it XMM-Newton} observations. In table \ref{table:2} we report the best-fit values of the parameters. The value of $T_{\rm{in}}$ turns out to be $\sim 0.3$ keV, compatible with those on the literature, for both the viewing geometries, which also show indistinguishable spectral results. Considering the results with $B=8\times10^{12}$ G, in the following, we took into account only the case with this value of the magnetic field strength.  

\begin{table*}
\caption{Best fit parameters for M51 ULX-7 ($B=8\times10^{12}$ G) }
\label{table:2}      
\centering                          
\renewcommand\arraystretch{1.2}
\begin{tabular}{ c c c c c c c c c  llllllll }     
 \hline \hline
& \multicolumn{7}{c}{$\chi=20^\circ$, $\xi=10^\circ$} \\
\multicolumn{1}{c}{Observation ID} & N\tablefootmark{a} ($10^{23}$) & K\tablefootmark{b} ($10^{-6}$) & $\alpha$\tablefootmark{b} & $\beta$\tablefootmark{b} & J\tablefootmark{c} ($10^{-3}$) & $T_\mathrm{in}$\tablefootmark{d} & $\chi^{2}$/dof \\
 & ($\rm{keV/cm^2/s}$) & ($\rm{keV/cm^2/s}$) &  & (keV) & ($\rm{keV/cm^2/s}$) &  (keV) &   \\
 \cmidrule(r){2-8} \\
 0824450901 & 5.8 $\pm$ 0.8  & 1.73 $\pm$ 1.73   &  -5.1 $\pm$ 1.1  &  1.3 $\pm$ 0.3  &  3.3 $\pm$ 0.2 & 0.32  & 426/384\\ 
 0830191501 & 3.7 $\pm$ 0.2  & 0.094 $\pm$ 0.093   &  -8.2 $\pm$ 1.1  &  0.9 $\pm$ 0.1  &  4.2 $\pm$ 0.2 & 0.32  & 445/380\\ 
 0830191601 & 9.8 $\pm$ 7.6  & 39.3 $\pm$ 31.9   &   -1.6 $\pm$ 0.6   &  5.1 $\pm$ 1.9  &  3.6 $\pm$ 1.0 & 0.28  & 423/380\\
 \\
 & \multicolumn{7}{c}{$\chi=60^\circ$, $\xi=10^\circ$} \\
\multicolumn{1}{c}{Observation ID} & N\tablefootmark{a} ($10^{23}$) & K\tablefootmark{b} ($10^{-6}$) & $\alpha$\tablefootmark{b} & $\beta$\tablefootmark{b} & J\tablefootmark{c} ($10^{-3}$) & $T_\mathrm{in}$\tablefootmark{d}  &  $\chi^{2}$/dof \\
 & ($\rm{keV/cm^2/s}$) & ($\rm{keV/cm^2/s}$) &  & (keV) & ($\rm{keV/cm^2/s}$) &  (keV) & \\
 \cmidrule(r){2-8} \\
 0824450901 & 10.1 $\pm$ 1.2   & 1.26 $\pm$ 1.28     &  -5.3 $\pm$ 1.1   &  1.3 $\pm$ 0.3  &  3.5 $\pm$ 0.2 & 0.32  & 431/384\\ 
 0830191501 & 6.6  $\pm$ 0.3   & 0.046 $\pm$ 0.049   &  -8.9 $\pm$ 1.2   &  0.8 $\pm$ 0.1  &  4.5 $\pm$ 0.2 & 0.32  & 448/380\\ 
 0830191601 & 19.2 $\pm$ 13.2  & 43.5 $\pm$ 26.9     &   -1.5 $\pm$ 0.4  &  5.6 $\pm$ 1.6  &  3.5 $\pm$ 0.9 & 0.30  & 420/380\\
 \hline
\end{tabular}
\tablefoot{\\
\tablefoottext{a}{{\sc torusdisk} norm}\\
\tablefoottext{b}{ {\sc cutoffpl}: norm, photon index, and e-folding energy of exponential cut-off}\\
\tablefoottext{c}{ {\sc bbody} norm} \\
\tablefoottext{d}{Temperature at the inner radius of the accretion disk}}
\end{table*}

\subsubsection{NGC 7793 P13}
\begin{figure*}
\centering
    \includegraphics[scale=0.5]{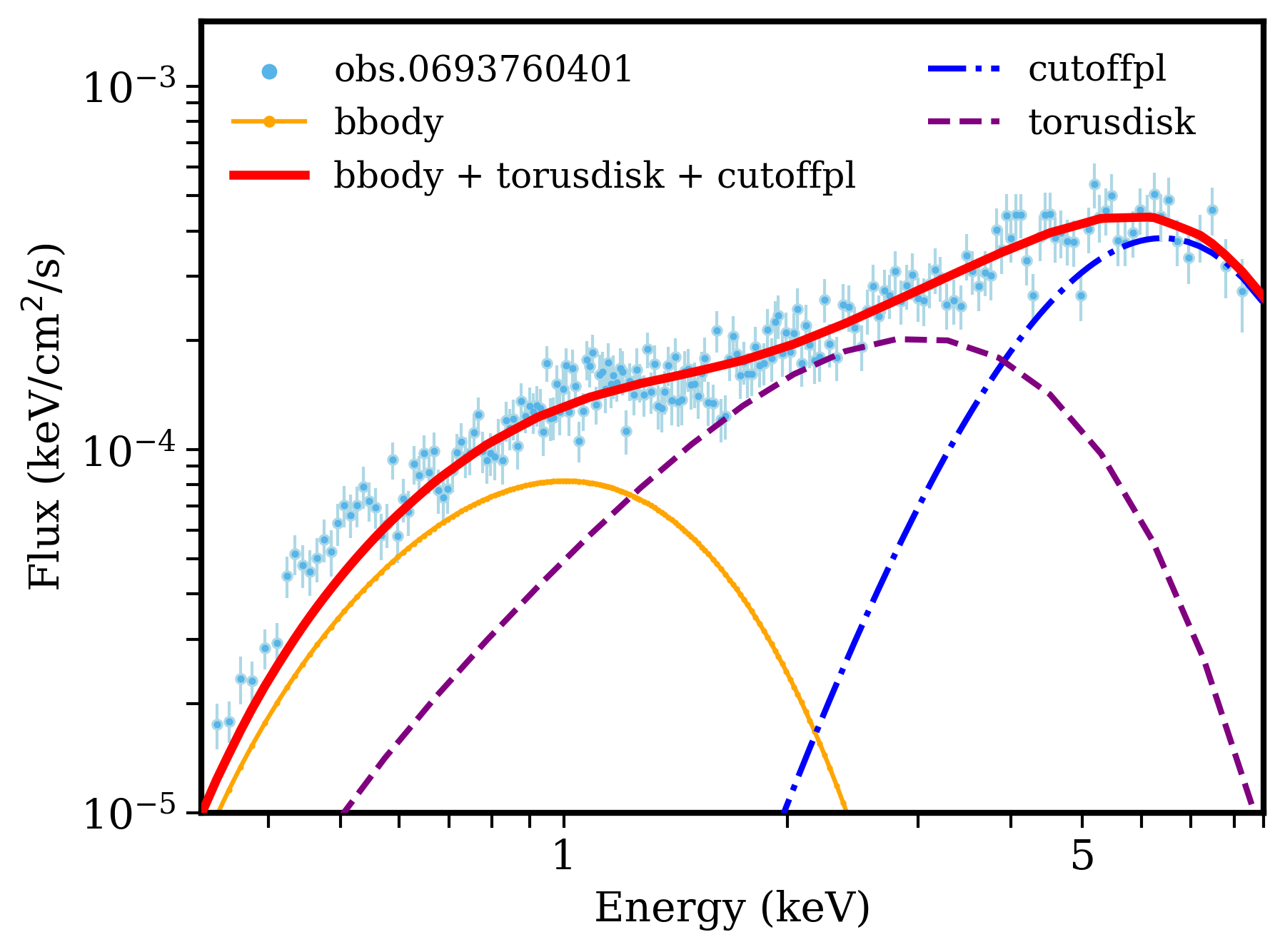}
    \includegraphics[scale=0.5]{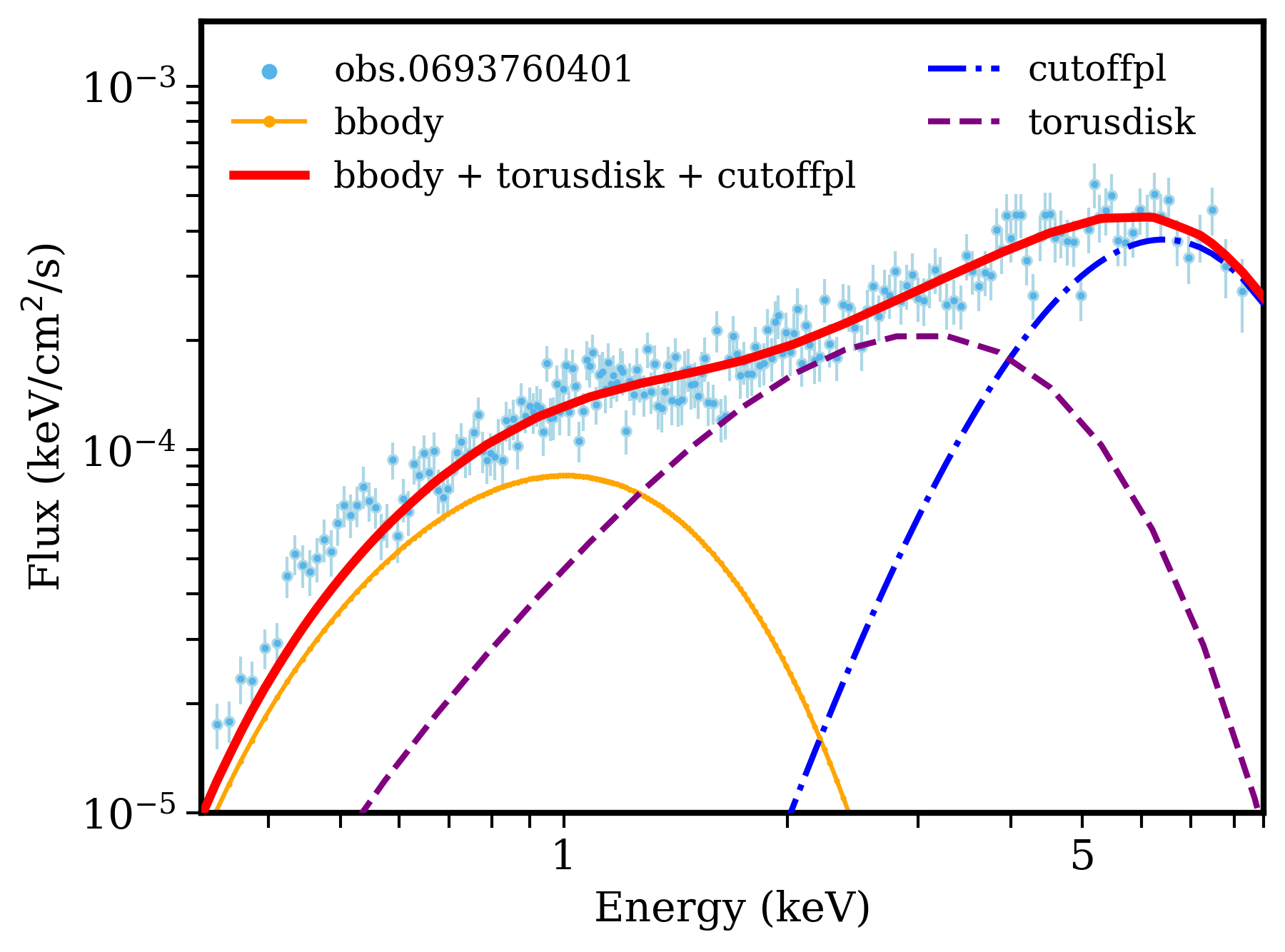}
    \includegraphics[scale=0.5]{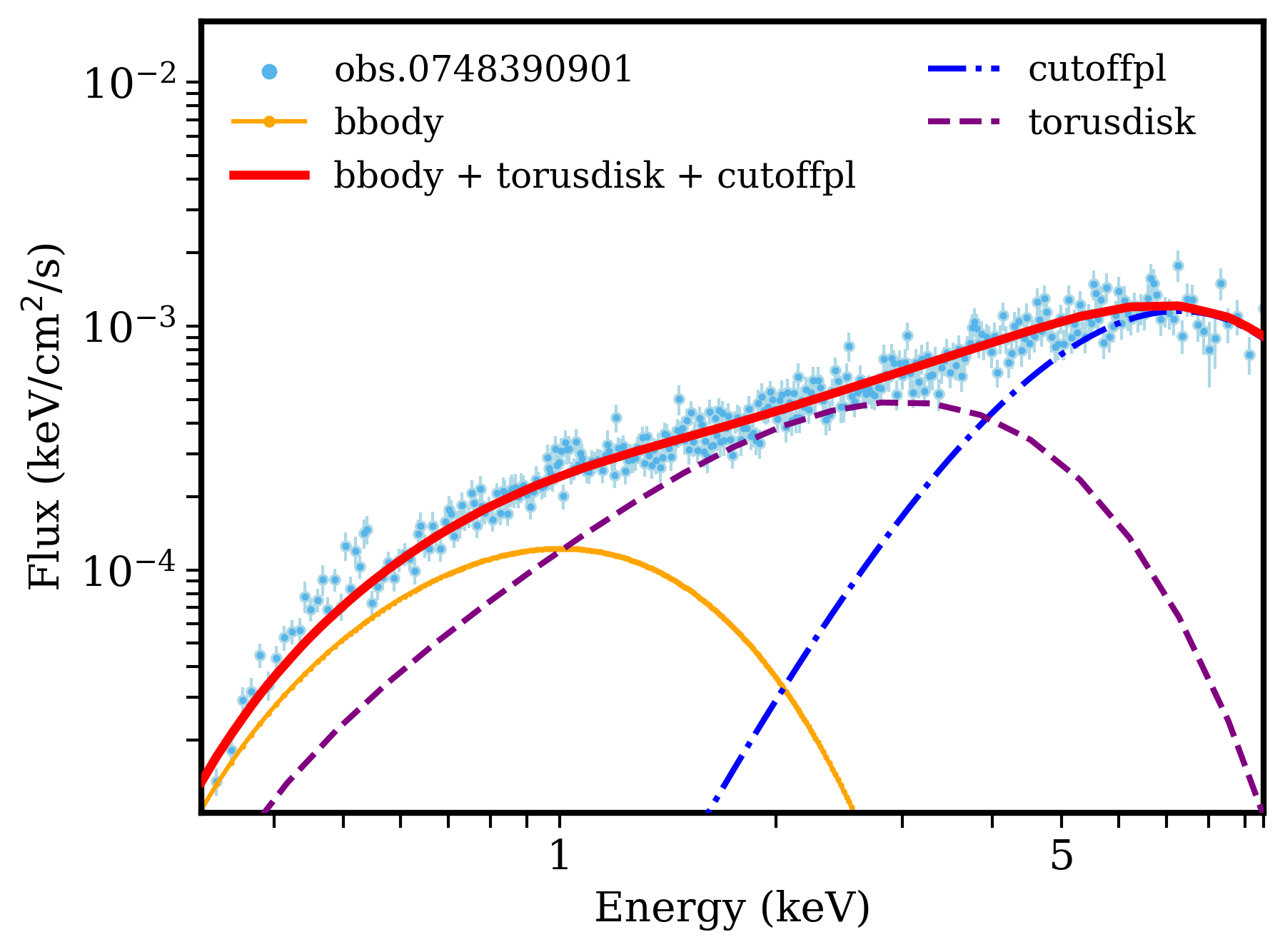}
    \includegraphics[scale=0.5]{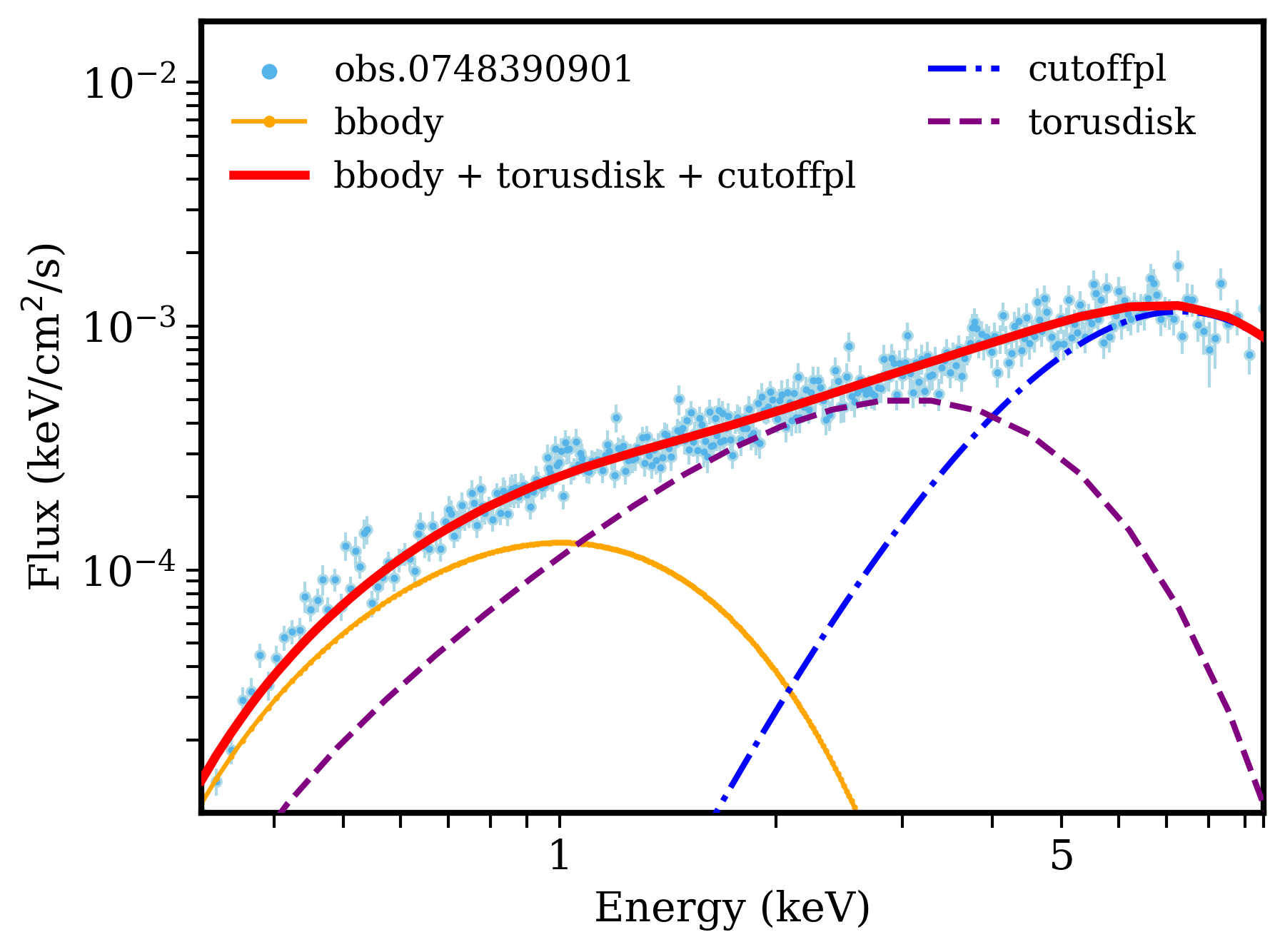}  
    \caption{X-ray spectrum of NGC 7793 P13 obtained with {\it XMM-Newton} (observations: 0693760401 on the top, 0748390901 on the bottom). The spectrum is fitted with a multicomponent model. Two components are phenomenological ({\sc bbody}, {\sc cutoffpl}), and one is computed with the model {(\sc torusdisk)}. Panels on the top (bottom) from left to right are comparisons with geometries $\chi=40^\circ$ and $\chi=60^\circ$, respectively.}
\label{fit_P13}
\end{figure*}

We adopted the same approach for the spectral fits on NGC 7793 P13, using the geometries $\chi=60^\circ$ $\xi=10^\circ$ and $\chi=40^\circ$ $\xi=10^\circ$ derived form the pulse profiles analysis. From the X-ray spectral analyses of \citet{walton2018super}, the spectrum of this source can be fitted with a model made by three components, two of which are thermal and dominate below 10 keV. They found that the temperature of the coolest thermal component spans from $\sim 0.3$ to $0.5$ keV, and that of the hottest thermal component from $\sim 1$ to $1.5$ keV. We used the same multicomponent model described above (bbody + torusdisk model + cutoff power law), correcting for absorption ({\sc wabs}) with a fixed column density, $N_{H}=\sim 3\times10^{20}\,\,\rm{cm^{-2}}$, and temperature of the {\sc bbody} component, 0.25 keV. We then varied $T_\mathrm{in}$ between $0.25$-- $0.5$ keV in 16 bins. We show the results in figure \ref{fit_P13} and we report the values of the parameters in table \ref{table3}. We find $T_{\rm{in}}\sim 0.3$ keV, in agreement with the literature, and also in this case the spectra of the two geometries of view do not show differences. The results also suggest that $B=4\times 10^{12}$ G can reproduce the spectrum well.

\begin{table*}
\caption{Best fit parameters for NGC 7793 P13 }
\label{table3}      
\centering                          
\renewcommand\arraystretch{1.2}
\begin{tabular}{ c c c c c c c c c llllllll }     
 \hline \hline
& \multicolumn{7}{c}{$\chi=40^\circ$, $\xi=10^\circ$} \\
\multicolumn{1}{c}{Observation ID} & N\tablefootmark{a} ($10^{23}$) & K\tablefootmark{b} ($10^{-6}$) & $\alpha$\tablefootmark{b} & $\beta$\tablefootmark{b} & J\tablefootmark{c} ($10^{-3}$) & $T_\mathrm{in}$\tablefootmark{d} & $\chi^{2}$/dof \\
 & ($\rm{keV/cm^2/s}$) & ($\rm{keV/cm^2/s}$) &  & (keV) & ($\rm{keV/cm^2/s}$) &  (keV) & & \\
 \cmidrule(r){2-9} \\
 0693760401 & 3.5 $\pm$ 0.2  &  0.6 $\pm$ 0.4   &  -7.5 $\pm$ 0.8  &  0.8 $\pm$ 0.1  &  4.5 $\pm$ 0.3 & 0.32 & 329/197\\ 
 0748390901 & 1.46 $\pm$ 0.08  & 2.0 $\pm$ 1.1   &  -6.5 $\pm$ 0.6  &  1.1 $\pm$ 0.1  &  6.8 $\pm$ 0.7 & 0.32 & 455/337\\ 
 \\
 & \multicolumn{7}{c}{$\chi=60^\circ$, $\xi=10^\circ$} \\
\multicolumn{1}{c}{Observation ID} & N\tablefootmark{a} ($10^{23}$) & K\tablefootmark{b} ($10^{-6}$) & $\alpha$\tablefootmark{b} & $\beta$\tablefootmark{b} & J\tablefootmark{c} ($10^{-3}$) & $T_\mathrm{in}$\tablefootmark{d} & $\chi^{2}$/dof \\
 & ($\rm{keV/cm^2/s}$) & ($\rm{keV/cm^2/s}$) &  & (keV) & ($\rm{keV/cm^2/s}$) &  (keV) & & \\
 \cmidrule(r){2-9} \\
 0693760401 & 4.8 $\pm$ 0.3  & 0.50 $\pm$ 0.35   &  -7.7 $\pm$ 0.9  &  0.83 $\pm$ 0.09  &  4.7 $\pm$ 0.3 & 0.32 & 334/197\\ 
 0748390901 & 2.0 $\pm$ 0.1  & 1.6 $\pm$ 0.9   &  -6.6 $\pm$ 0.6  &  1.1 $\pm$ 0.1  &  7.1 $\pm$ 0.7 & 0.32 &457/337\\ 
 \hline
\end{tabular}
\tablefoot{\\
\tablefoottext{a}{{\sc torusdisk} norm}\\
\tablefoottext{b}{ {\sc cutoffpl}: norm, photon index, and e-folding energy of exponential cut-off}\\
\tablefoottext{c}{ {\sc bbody} norm} \\
\tablefoottext{d}{Temperature at the inner radius of the accretion disk}}
\end{table*}

\subsection{Polarization} 
\label{subsec:polarization properties}
In simulating the polarization properties of the two PULXs, we considered fields with strengths of $4\times10^{12}$ G (NGC 7793 P13) and $8\times10^{12}$ G (M51 ULX-7) at the poles. For photons escaping from the adiabatic region, we fixed the intrinsic polarization fraction, $\Pi_\mathrm{L}$, at $60\%$ in the X-mode. We did not computed this value self-consistently since this would have required to solve the full radiation transport problem, which is definitely outside the scope of the present work. Nevertheless, in the presence of strong magnetic fields ($\gtrsim 10^{13}$ G) X-mode photons are expected to dominate the emission, so our assumption is not unreasonable. The effects of changing $\Pi_{\rm{L}}$ will be discussed further on. We then analyzed the variation of the polarization degree and polarization angle with the geometry of view. In the case of M51 ULX-7 we used $\chi=$ $60^\circ$, $40^\circ$ and $20^\circ$, while for NGC 7793 P13 $\chi=$ $60^\circ$ and $40^\circ$. 

\subsubsection{M51 ULX-7}
\label{pol_M51}
\begin{figure*}
\centering
    \includegraphics[scale=0.5]{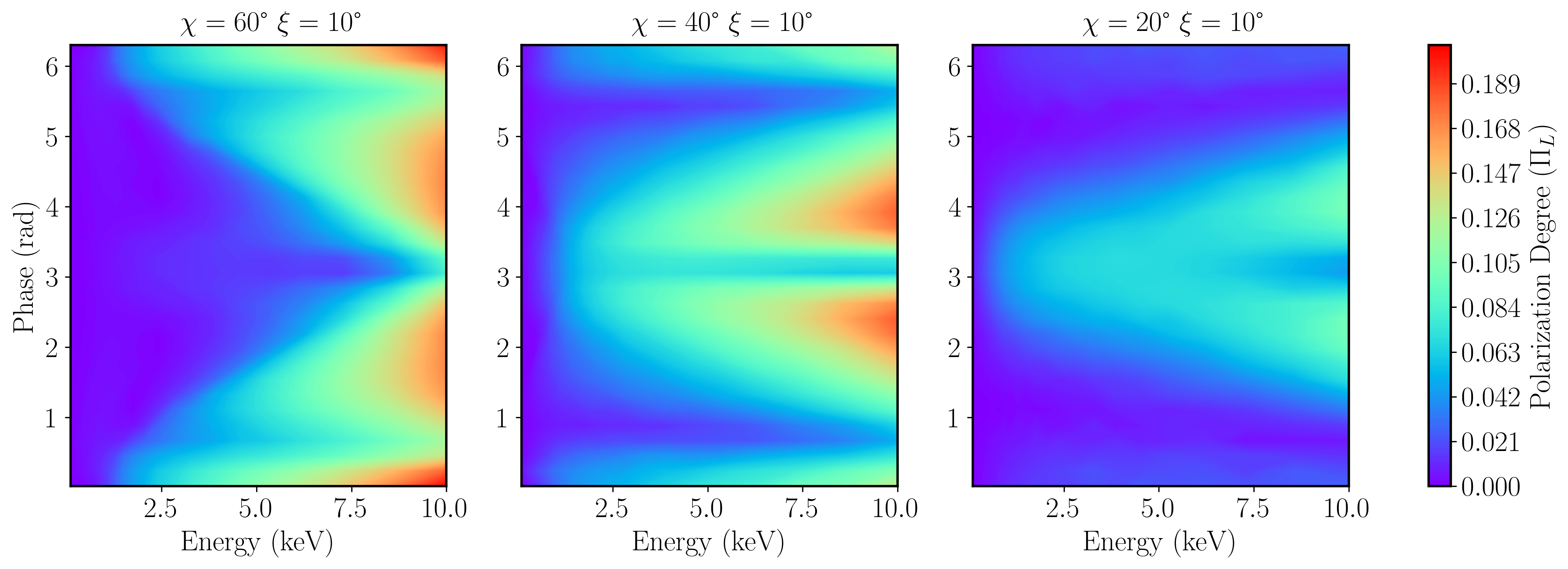}
    \includegraphics[scale=0.5]{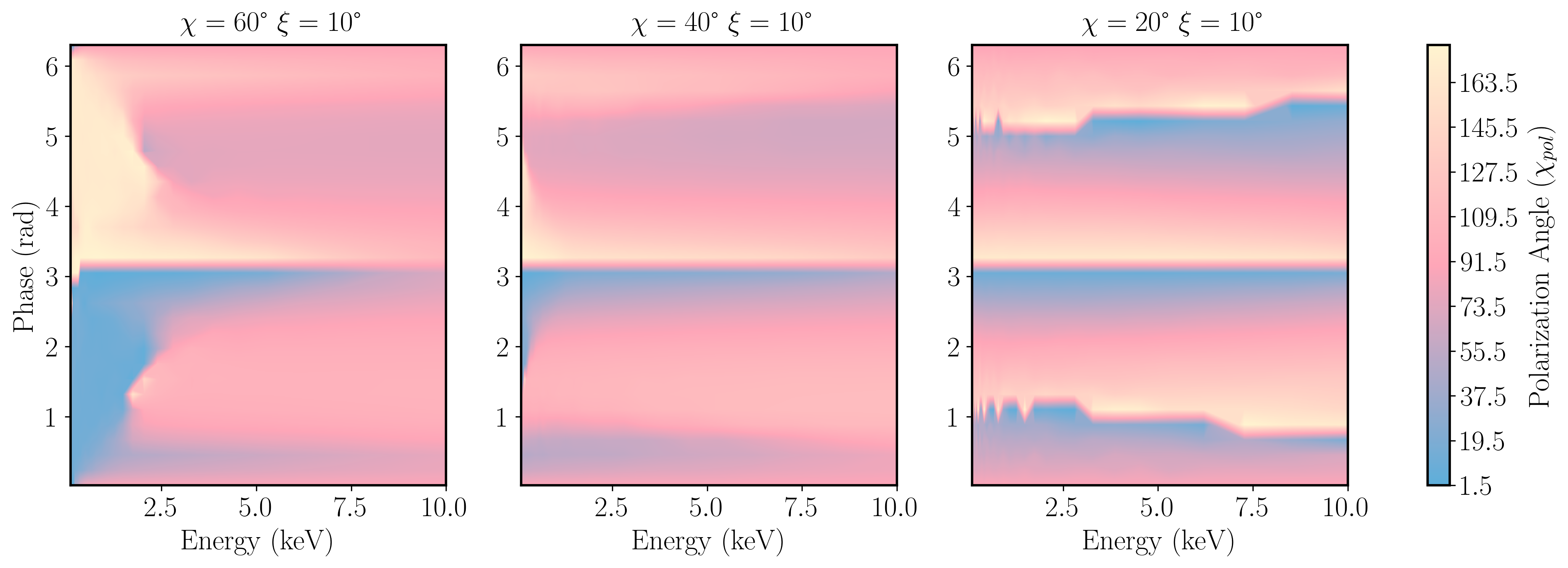}
    \caption{PD and PA variation with energy and phase for different geometries of view. From left to right $\chi=$ $60^\circ$, $40^\circ$, and $20^\circ$.}
\label{PD_M51_G_variations}
\end{figure*}
We show in figure \ref{PD_M51_G_variations} the energy- and phase-dependent PD, for three different viewing geometries.
In each panel, the polarization degree increases with the energy and shows an oscillating behavior with the phase, that changes going from $\chi=20^\circ$ to $\chi=60^\circ$. The former is a consequence of the strong magnetic field considered in the model, while the latter is a consequence of the rotation of the source combined with the geometry of view. 

Fixing energy and phase but varying the angle $\chi$ from $20^\circ$ to $60^\circ$ the value of the polarization degree increases. This effect could be caused by the geometrical depolarization. The polarization angle oscillates with the phase remaining almost constant with the energy. As for the polarization degree, the variation with the phase of the polarization angle depends on the viewing geometry.

\subsubsection{NGC 7793 P13}
\begin{figure*}
\centering
    \includegraphics[scale=0.45]{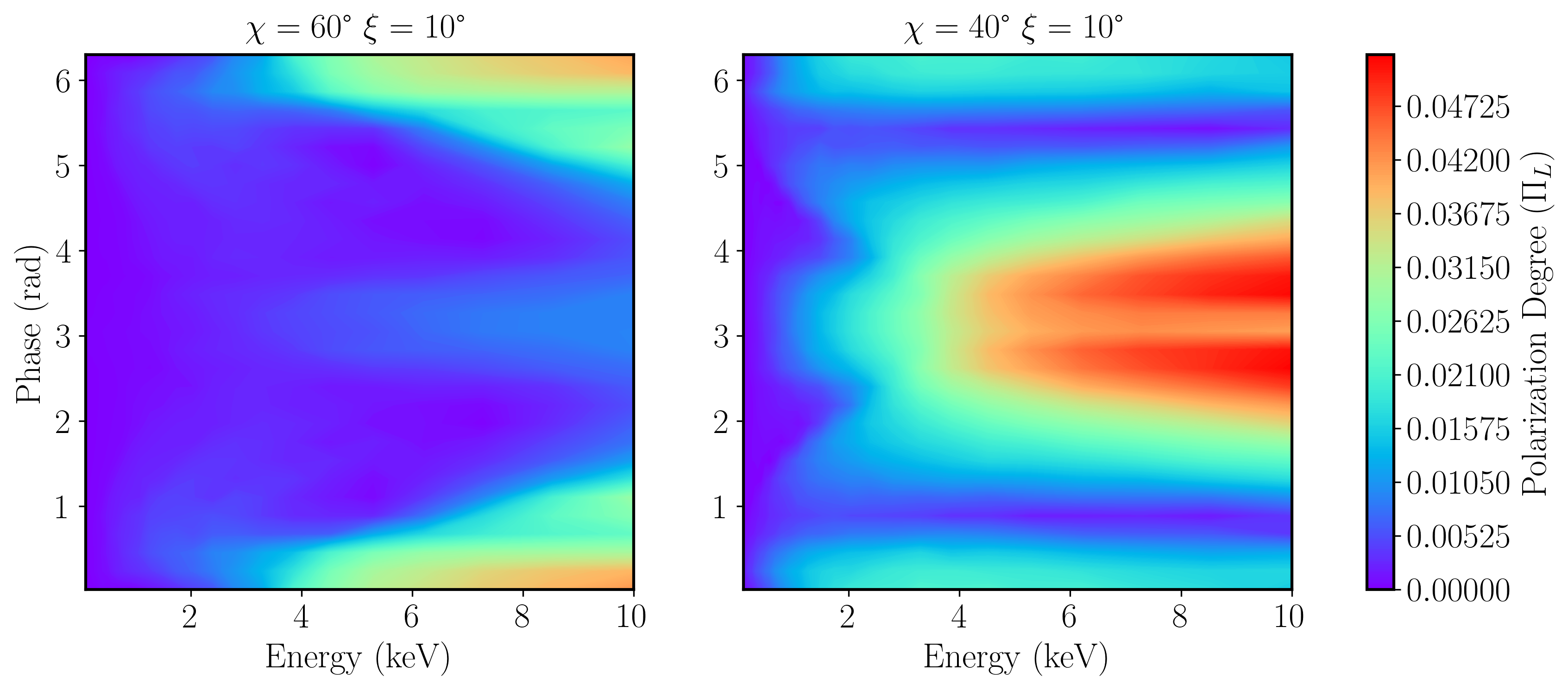}
    \includegraphics[scale=0.45]{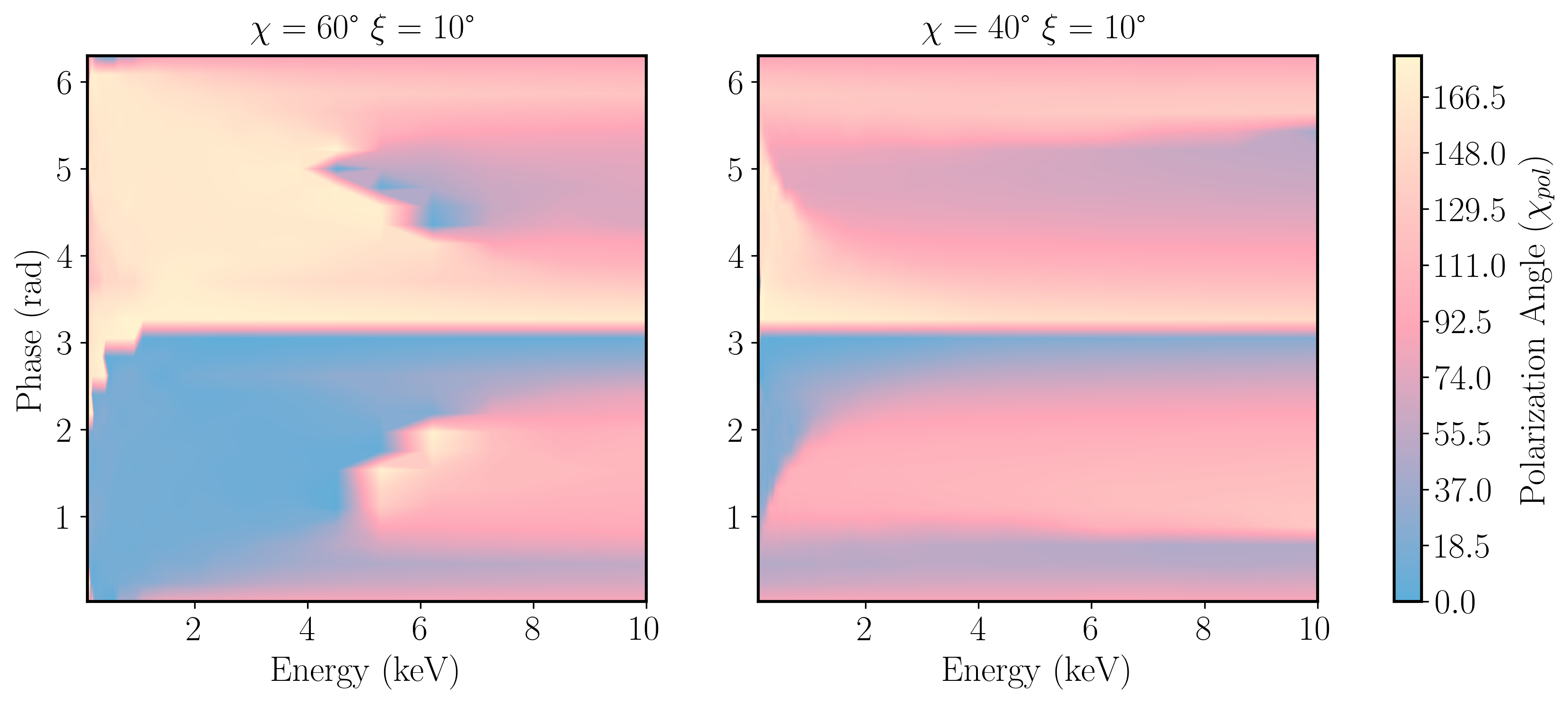}
    \caption{PD and PA variation with energy and phase for different geometries of view. From left to right $\chi=$ $60^\circ$ and $40^\circ$.}
\label{PD_P13_G_variations}
\end{figure*}

In figure \ref{PD_P13_G_variations} we show the polarization results for NGC 7793 P13. 
The maximum polarization degree is lower than that for M51 ULX-7. Fixing energy and phase, it is higher for the geometry of view with $\chi=40^\circ$ than that with $\chi=60^\circ$. As for M51 ULX-7, the polarization degree oscillates with the phase and increases with the energy, while the polarization angle oscillates with the phase remaining quite constant with the energy for low values of $\chi$.

\subsubsection{Polarization results with different $\Pi_{\rm{L}}$}
We previously mentioned that the value of $\Pi_{\rm{L}}$ is not computed, but it is fixed considering the magnetic field strength of the source. We report here the results for the polarization degree varying $\Pi_{\rm{L}}$ and fixing the geometry of view with $\chi=60^\circ$ and $\xi=10^\circ$, for both M51 ULX-7 and NGC 7793 P13. The results show that a variation in the intrinsic polarization degree only leads to an overall variation in the value of the PD, without changing its behavior with energy and phase. This reinforces the hypothesis that, even if the intrinsic polarization value assumed in the model may be different from the one computed self-consistently, we can still distinguish the geometry of view through the energy and phase variations of PD, since they do not depend on the value of $\Pi_{\rm{L}}$, but only on the geometry of view and magnetic field of the source.
\begin{figure*}
\centering
\includegraphics[scale=0.5]{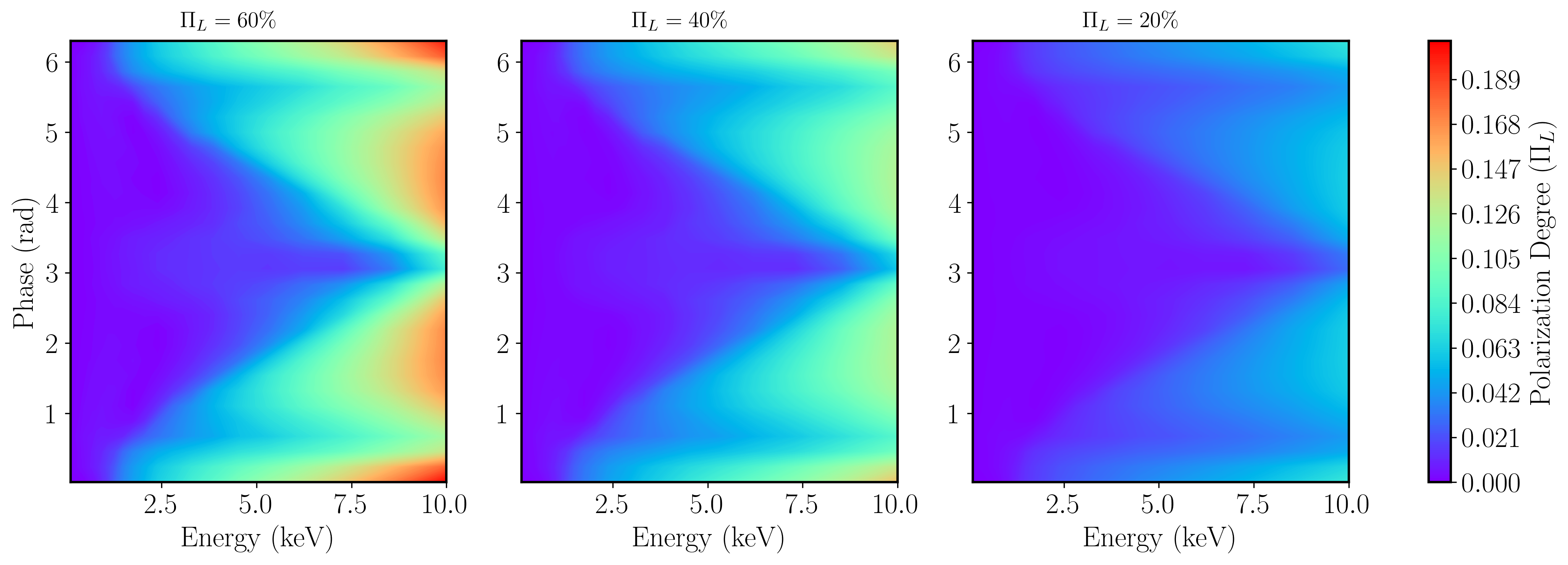}
\caption{Polarization degree variation for different values of $\Pi_{\rm{L}}$, for M51 ULX-7 with $B=8\times10^{12}$ G and geometry of view $\chi=60^\circ$ $\xi={10^\circ}$.}
\label{M51_PD_VarieN}
\end{figure*}

\begin{figure*}
\centering
\includegraphics[scale=0.5]{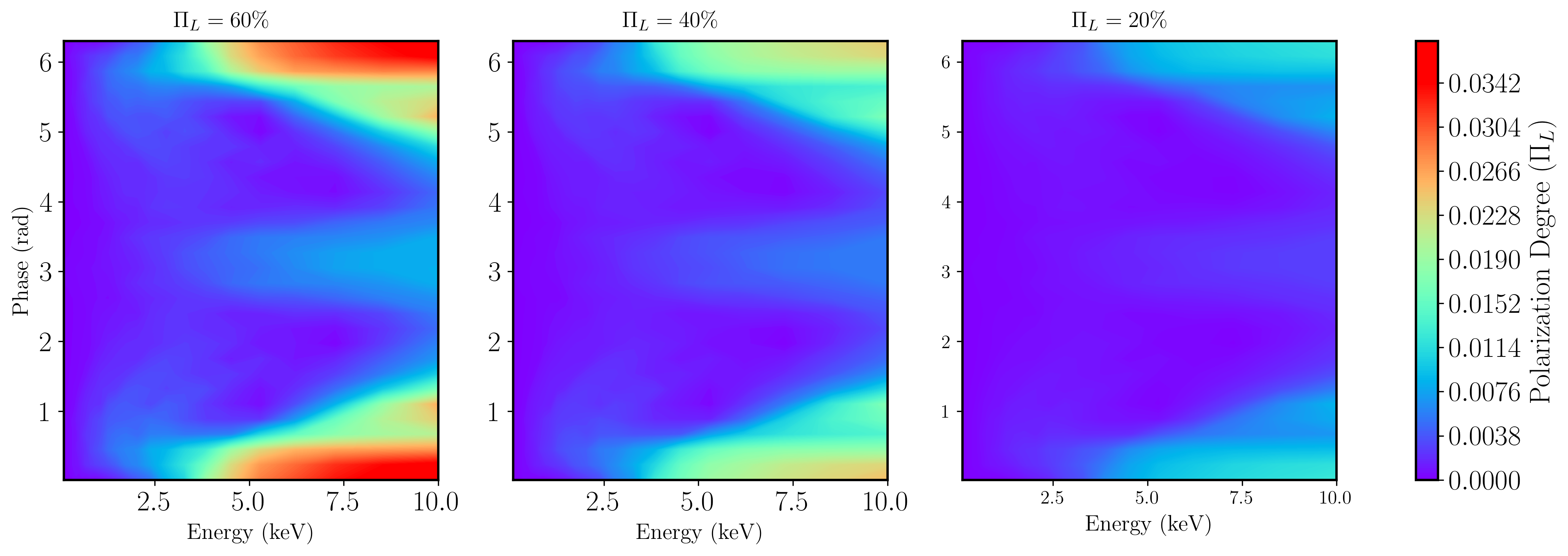}
\caption{Polarization degree variation for different values of $\Pi_{\rm{L}}$, for NGC 7793 P13 with $B=4\times10^{12}$ G and geometry of view $\chi=60^\circ$ $\xi=10^\circ$.}
\label{P13_PD_VarieN}
\end{figure*}

\section{Discussion}
\label{sec:discussions}
This paper aims to constrain the geometry of view, the magnetic field strength, and thermal properties of pulsating ULXs, reproducing the thermal radiation emitted by a highly magnetized neutron star accreting at a super-Eddington rate and taking into account the emission from an accretion disk and an optically thick envelope surrounding the magnetosphere. The model reproduces the light curves and the spectra of PULXs, for the latter by adding one or two phenomenological components that are not self-consistently included in the model: a blackbody component to model the excess of flux at low energies, probably produced by optically thick winds from an intermediate region of the disk; a cut-off power law produced by an accretion column. In addition to the spectral properties, we also incorporated the polarization observables of the emitted radiation, focusing our analyses on the polarization degree. We discuss our results in the following.

\subsection{M51 ULX-7}
\subsubsection{Light curves and spectra}
We started comparing the light curves because the variation of the flux with the phase, and the corresponding pulsed fraction, are highly dependent on the viewing geometry, allowing us to constrain it. We considered light curves in two energy bands, $2$--$3$ keV and $3$--$4$ keV, where we expect our model, in particular the torus emission, to dominate. Being the emission from the accretion column, which contributes to the PF at high energies, not included in our model, it would be impossible to compute the PF self-consistently. For M51 ULX-7 we did not find a unique solution for the viewing geometry but we were able to derive a lower and upper limit, $\chi=20^\circ$, $\xi=10^\circ$ and $\chi=60^\circ$, $\xi=10^\circ$, respectively. This is not surprising since the source is very variable and its PF does not increase monotonically with the energy, as shown in Figure 3 in \cite{brightman2022evolution}. The reasons behind the source variability are not well known. They could be related to the fact that the source entered a propeller stage, or because of a variation in the mass transfer rate, which is unstable.

We used both the geometries to perform the spectral comparison, adding to the torusdiks multicolor blackbody the phenomenological components to fit the entire spectrum. We considered two representative values for the magnetic field, $B=10^{12}$ G and $8\times10^{12}$ G, respectively, which brackets the true value. For each of them, we computed the Alfvén radius $\ra$ and the torus temperature. We also fixed the disk outer radius, $R_{\rm{disk}}=200\,\rm{R_{\rm{NS}}}$, and varied the temperature at the inner radius of the disk between 0.25 and 0.5 keV. 

For a magnetic field $B=10^{12}$ G we were able to fit the spectrum with two thermal components: a blackbody at lower energies and our torusdik model at higher energies (figure \ref{M51_fit_B1}), similar to what was done by \cite{castillo2020discovery}. In the fitting procedure, we left free to vary the normalization of the torusdisk model and the column density $N_{\rm{H}}$, while we fixed at 0.25 keV the {\sc bbody} temperature, and we let the temperature $T_{\rm{in}}$ to vary between $0.25$ and $0.5$ keV, but fixing it for each simulated spectrum. We did not fix the parameters of the phenomenological {\sc bbody} component to the values inferred by \cite{castillo2020discovery}, because our multicolor blackbody is different. The two thermal component models of \cite{castillo2020discovery} fit well the spectrum, with a higher column density of $5.9\times10^{20}\rm{cm^{-2}}$, and a higher value for the temperature and radius of the hotter thermal component ($1.5$ keV and $97\, \rm{R_{\rm{NS}}}$, respectively), while $T_{\rm{in}}$ is compatible within the uncertainties with our value. On the other hand, our physically motivated model (blackbody + torusdisk) does not provide a satisfactory fit to the data (figure \ref{M51_fit_B1}) because of residuals between $\sim 1.5$ and $3.0$ keV. The reason is that in our torusdisk multicolor blackbody, the energy range of the torus temperature varies depending on the choice of the magnetic field strength and cannot be adjusted in the fit. In addition, \cite{castillo2020discovery} also adopted the absorption model {\sc tbabs}, an updated version of {\sc wabs}, and this can explain the higher $N_{\rm{H}}$.

For a magnetic field $B=8\times10^{12}$ G, we fitted the data with the superposition of three components: a phenomenological blackbody, the torusdisk model, and a phenomenological cutoff power law, which represents the non-thermal emission from the accretion column. With the three component model, the spectrum is reproduced with great accuracy (Figure \ref{comparison_spectra_M51}). A similar 3-component spectral analysis was also performed by \cite{brightman2022evolution}. They fitted the data using the same blackbody and cutoff power law components, and a phenomenological multicolor blackbody at intermediate energies, well reproducing the whole spectrum.
For the same reasons repeated above we did not use their best-fit parameters for the blackbody and the cutoff power law, but we left them free to vary (apart from those that we fixed to ensure the convergence of the fit: the {\sc bbody} temperature and the total column density). The $T_{\rm{in}}$ is $\sim 0.36$ keV, compatible with our value, while the column density $N_{\rm{H}}$ is lower than that found by \cite{castillo2020discovery}, $N_{\rm{H}}\sim 3.3\times10^{20}\pm 0.7 \,\rm{cm^{-2}}$, and closer to our value (the residual difference being again possibly related to the different absorption model). For the cutoff power law parameters, \cite{brightman2022evolution} fixed them using the values found by \cite{walton2018super}, $E_{\rm{cut}} (\rm{here}\, \beta) \sim 8.1$ keV and $\Gamma (\rm{here}\,\alpha) \sim 0.8$, that are different from our results. 

From spectral comparison we obtained good constraints on the temperature $T_{\rm{in}}$ and on the magnetic field strength $B$, but no additional constraints on the geometry of view. The latter can be further investigated through polarization analyses.

\subsubsection{Polarization}

As shown in Sect. \ref{pol_M51} the polarization degree increases with the energy (figure \ref{PD_M51_G_variations}). This is a consequence of the strong magnetic field considered. 
For $B=8\times10^{12}$ G, high energy photons are emitted from the torus inside the adiabatic region. This explains the oscillating behavior of the polarization degree with the phase, which is simply a consequence of the rotation of the source combined with the geometry of view, that changes the emitting regions in view from which the polarized radiation comes. For regions closer to the equator PD is low because photons have lower energies (the torus temperature is lower) and are emitted outside the adiabatic region.

Fixing the energy and the phase, the polarization degree increases with the angle $\chi$ (fig. \ref{PD_M51_G_variations}). This is because of the geometrical depolarization effect. To understand this point, let us consider two limiting cases: a pole-on configuration ($\chi=\xi$) and an equator-on configuration ($\chi-\xi=90^\circ$), with respect to the magnetic axis. In the first case, there is no preferential direction of the magnetic field vector, since the field assumes a radial configuration. Therefore photons escaping from those regions (linearly polarized parallel to or perpendicular to the plane of the vectors $\vec{k}$ and $\vec{B}$) will have an electric field with a radial configuration, which means to have opposite polarization contribution, so a null total polarization. 
In the second case, there is a preferential direction of the magnetic field vector, so summing the polarization contributions we have a non-zero polarization degree. For the $\chi=20^\circ$, $\xi=10^\circ$ geometry, we are closer to a pole-on configuration, while increasing $\chi$ up to $60^\circ$ we move towards an equator-on configuration.

It is also worth noting that even if we cannot distinguish the viewing geometry from the X-ray spectrum, polarization measurements may allow us to do it. Indeed, the behavior of the polarization degree remains the same even changing the intrinsic polarization fraction $\Pi_{\rm{L}}$.

The polarization angle shows an oscillating behavior with the phase, going from $90^\circ$ to $0^\circ$ or $180^\circ$, as expected from a rotating vector model, while remaining almost constant with the energy. The oscillations are caused by the geometry of view combined with the rotation of the star. Because the part in view of the emitting components varies from regions emitting polarized radiation (X-mode photons), where we expect the polarization angle to oscillate around $90^\circ$, to regions emitting mainly unpolarized radiation, that give a PA of $0^\circ$ or $180^\circ$, if we recall how PA is defined (equation (\ref{pol_frac_and_pol_ang})), and that the values of $U$ and $Q$, in those regions, are close to zero.

\subsection{NGC 7793 P13}

\subsubsection{Light curves and spectra}
We repeated the previous analysis on NGC 7793 P13, starting from the light curve comparisons. We found two limiting viewing geometries (this time with a smaller range of $\chi$): $\chi=40^\circ$, $\xi=10^\circ$, and $\chi=60^\circ$, $\xi=10^\circ$. The pulsed fraction of NGC 7793 P13 presents a more regular behavior, increasing quasi-monotonically with energy \cite[][]{israel2017discovery}. We used both geometries to perform the spectral fits, considering a magnetic field strength of $4\times10^{12}$ G (computing the torus temperature, the Alfvén radius, table \ref{tabella_temp_P13}, and range of $T_{\rm{in}}$ as before), and setting the disk outer radius $R_{\rm{disk}}=200\,\rm{R_{NS}}$. The spectrum is well reproduced by the superposition of a blackbody, the torusdisk model, and a cutoff power law. \cite{israel2017discovery} performed a phenomenological fit using an absorbed power law with a high energy cut-off and a multicolor blackbody ({\sc phabs[highecut*(powerlaw+bbodyrad)]}), obtaining the following parameters: $N_{\rm{H}}=9.6\times10^{20}\,\rm{cm^{-2}}$, $\Gamma (\rm{here\, \alpha}) \sim 1.2$, $E_{\rm{cut}}(\rm{here\,\beta}) \sim6$ keV, $kT_{\rm{BB}} \sim 0.2$ keV. Despite some differences in the fitting parameters, both models agree on the type of emission at high energies, well described by a power law, probably produced by an accretion column, while the emission in the soft X-ray energy band is thermal and well described by a multicolor blackbody.

\subsubsection{Polarization}
As for M51 ULX-7, also in NGC 7793 P13 for a given geometry of view the polarization degree increases with energy, because of the strong magnetic field, and it oscillates with phase. The maximum polarization degree is around $4.7\%$, lower than that of M51 ULX-7 ($\sim 19\%$), using the lower magnetic field strength, $B=4\times10^{12}$ G. For such magnetic field we expect the radiation to be poorly polarized (in the X-mode), expecting less variation of PD between more energetic and less energetic radiation (figure \ref{PD_P13_G_variations}, top right). We observe oscillations with the phase still related to the rotation of the source combined with the geometry of view.
The reason why this time the polarization degree decreases with $\chi$ depends on the different magnetic field strength used, on the geometry of view, and on how the geometrical depolarization effect dominates over the vacuum birefringence and vice versa. Indeed, despite the favorable viewing angle ($\chi=60^\circ$), because of the depolarization effect mentioned above, the lower magnetic field strength reduces the polarized radiation. As a result, in regions near the equator the radiation is so weakly polarized that, even summing all the polarization contributions, we cannot achieve a higher polarization degree (figure \ref{PD_P13_G_variations}, top left).

Also for NGC 7793 P13, the behavior of PD may allow us to distinguish the geometry of view if observed polarization measurements are available.

The polarization angle oscillates with phase and remains constant with energy, but switching drastically between the highest ($\sim 180^\circ$) and lowest value ($\sim 1^\circ$) at lower energies for the geometry of view $\chi=60^\circ$ $\xi=10^\circ$. The reason is probably that the radiation is mainly unpolarized.

\section{Conclusions}
\label{sec:summary}
In this work, we presented a simplified model that reproduces the X-ray thermal emission of pulsating ultraluminous X-ray sources, in terms of emission from an accretion disk and an accretion envelope modeled as a torus confined by the magnetic field lines and extending up to the magnetospheric radius. We used and modified the ray-tracer code discussed in \cite{taverna2015polarization}, to reproduce spectra and light curves of PULXs in two different energy bands. The model also simulates the polarization degree and angle of the radiation emitted as a function of energy and phase.
We tested the model on two PULXs: M51 ULX-7 and NGC 7793 P13, using for both {\it XMM-Newton} observations. The aim was to test the validity of our model and derive information on the geometry of view, magnetic field, thermal and polarization properties of such sources. 

In the model we considered the emission from a highly magnetized neutron star in an accreting binary system with a mass of $1.4\,{M_{\odot}}$, radius $10$ km, and magnetic field strength between $10^{12}$ and $\sim 10^{13}$ G. The accreting material originates from the donor star and proceeds through a geometrically-thin accretion disk up to the Alfvén radius, where particles are funneled along the magnetic field lines, which follow a dipolar field topology, towards the magnetic poles of the star. 
Each point of the torus and the disk emits like a blackbody, with a local temperature that depends on the magnetic colatitude $\theta$, if the point belongs to the torus, or on the radial distance if the point belongs to the disk. The emission of the whole source is then a complex multicolor blackbody. We considered as well polarization observables, taking as polarized only the radiation coming from below the adiabatic radius $\rpl$, while that coming from the disk and the part of the torus outside the region of adiabatic propagation is unpolarized. We computed the flux and the polarization observables for different viewing geometries ($\chi$, $\xi$) of each source. 

We made direct comparisons of the light curves considering the $2$–$3$ keV and $3$–$4$ keV energy bands. We compared each observed profile with six simulations having different geometries of view, and from these comparisons, we constrained the viewing geometry obtaining a lower and upper limit of the angle $\chi$, which was then used in the spectral analysis.

We compared spectra from different observations with those simulated with the model, varying the magnetic field strength, and using the temperature at the inner radius of the disk, $T_{\rm{in}}$, as a free parameter. For M51 ULX-7 we used three {\it XMM-Newton} observations (0824450901, 0830191501 and 0830191601), and compared the spectrum with a multicomponent model that combines a blackbody ({\sc bbody}, between $0.1-1.5$ keV), a multicolor blackbody (output of the model, between $1.5-5$ keV), and a cut-off power law ({\sc cutoffpl}, between $5-10$ keV), assuming a magnetic field strength $B=8\times10^{12}$ G. The best fitting internal disk temperature is $T_\mathrm{in}\approx0.3$ keV. We did the same for NGC 7793 P13, using the {\it XMM-Newton} observations 0693760401 and 0748390901, finding a $T_\mathrm{in}\approx0.32$ keV. 

For a given viewing geometry, the polarization degree increases with the energy and shows an oscillating behavior with the phase. The former is a consequence of the strong magnetic field considered in the model, while the oscillating behavior with the phase is a consequence of the rotation of the source combined with the geometry of view. Fixing the energy and phase but varying the angle $\chi$ we note that the value of the polarization degree increases. The variation of PD from one geometry of view to another could allow us to distinguish them if observed polarization measurements are available.
Therefore, the polarization analysis has proved to be crucial in determining the geometry of view of the source when the spectral one gives degenerate results.  

The model reproduces well the observed spectra of the two pulsating ULXs, particularly in the $1.5$--$5$ keV interval, where the contribution of the torus with the disk is dominant.

\begin{acknowledgements}
     SC is supported by the INAF Doctorate program. SC and LZ acknowledge financial support from the INAF Research Grant ``Uncovering the optical beat of the fastest magnetized neutron stars (FANS)". RTa and RTu acknowledge financial support from the Italian MUR through grant PRIN 2022LWPEXW. SC thanks E. Giunchi, B. Ambrosio and I. Salmaso for technical support. 
\end{acknowledgements}



%
  

%
%
\bibliographystyle{aa}
\bibliography{biblio}

%


\end{document}